# Application of a Pressure-Based OpenFOAM Solver for Rotating Detonation Engines

Keunjae Kwak[1], Hyoungwoo Kim[1], Je Ir Ryu[2,3], Dong-Hyuk Shin[1,*]

*Affiliations*

[1]Department of Aerospace Engineering, Korea Advanced Institute of Science and Technology, Daejeon, 34141, Republic of Korea

[2]Division of Engineering, New York University Abu Dhabi, United Arab Emirates

[3]Tandon School of Engineering, New York University, Brooklyn, NY, 11201, USA

[*] Correspondence: Dong-Hyuk Shin (donghyuk.shin@kaist.ac.kr)

# Abstract

This study presents an open-source, pressure-based simulation framework for rotating detonation engines (RDEs) using the *multicomponentFluid* solver in OpenFOAM v12. A two-dimensional adaptive mesh refinement (AMR) method is developed and coupled with dynamic load balancing (DLB) to reduce computational cost. The framework is validated against one-dimensional detonation-tube experiments, a detailed Zeldovich–von Neumann–Döring (ZND) solution, two-dimensional cellular-detonation data, and a two-dimensional RDE reference simulation. The calculated detonation velocities agreed with the Chapman–Jouguet (CJ) value within 0.1%, and the ZND induction length was resolved within 1% at grid spacings of 0.5 μm or smaller. The simulations also captured the principal cellular scale and characteristic RDE flow structures and performance. After accounting for detonation-front inclination and stream-tube expansion, the horizontal detonation velocity differed from the corrected CJ value by 1.27%. The AMR and AMR+DLB results differed from the uniform-grid predictions by no more than 1.07%. AMR alone provided a 1.5-fold speed-up, whereas AMR+DLB achieved a 4.2-fold speed-up, demonstrating the computational efficiency of the framework for the present two-dimensional RDE configuration.

# 1. Introduction

Aerospace propulsion systems based on constant-pressure (Brayton cycle) combustion, including turbofan engines and liquid rocket engines, have reached a high level of maturity, with thermal efficiency and specific impulse approaching fundamental thermodynamic and material limits [1]. Detonation-based combustion achieves a simultaneous pressure and temperature rise during the combustion process, and thermodynamic analyses have shown that the idealized detonation cycle can yield approximately 20% higher thermal efficiency than the Brayton cycle [2], translating to potential specific impulse gains on the order of 10% for rocket applications [3]. Among various detonation-based engine concepts, rotating detonation engines (RDEs) have emerged as particularly attractive due to their continuous operation, mechanical simplicity, and compact annular geometry that enables high thrust-to-weight ratios [4, 5].

In addition to rocket propulsion, RDEs are increasingly being investigated as pressure-gain combustors for gas-turbine-based propulsion and power-generation systems [6, 7]. Numerical simulations of such integrated RDE-turbine systems must accommodate a broad range of flow regimes, from shock-dominated detonation flow within the combustor to subsonic and transonic flow through downstream components. Because substantial portions of the integrated system operate in low-Mach regimes, pressure-based formulations, which are generally efficient and robust under such conditions, provide an attractive basis for a unified simulation framework. However, pressure-based formulations have been less commonly applied to shock-dominated detonation flows, and their ability to predict global RDE flow and performance quantities must therefore be carefully assessed. These considerations support using OpenFOAM's built-in pressure-based *multicomponentFluid* solver as the foundation of the present RDE simulation framework and as a first step toward fully integrated RDE-turbine simulations.

Detonation differs fundamentally from deflagration. It can be regarded as a reacting shock wave [8]. Reactants are compressed and heated by the leading shock during detonation [9]. Thermodynamic properties such as pressure and temperature rise sharply across a detonation wave. Moreover, detonation waves propagate at supersonic speeds on the order of $10^3$ m/s, whereas deflagration flame speeds are subsonic and on the order of 1 m/s.

These characteristics give rise to two advantages: higher thermal efficiency and compact engine configurations. Detonation-based combustion can potentially improve thermal efficiency in engine systems.

Wintenberger and Shepherd analyzed the thermodynamic cycle of propagating detonations and concluded that an idealized detonation-based cycle exhibits higher thermal efficiency than constant-volume or constant-pressure combustion processes for identical initial conditions ahead of the combustion wave [2]. Improved efficiency can reduce fuel consumption, which is closely linked to emissions. In addition, the high propagation speed allows more reactants to be consumed within a given time, substantially increasing the heat-release rate. Consequently, this characteristic enables compact combustor configurations without sacrificing the desired performance.

Several detonation-based engine concepts have been proposed: standing detonation engines, pulse detonation engines (PDEs), and rotating detonation engines (RDEs) [10]. Among these, this study focuses on RDEs. RDEs use a distinct mechanism to sustain detonation waves within the combustor. An RDE typically has an annular cylindrical geometry with a narrow channel. Reactants are injected from the head end, and one or more detonation waves propagate azimuthally through the channel while the combustion products exit from the opposite end.

RDEs offer several notable advantages because of this operating mechanism. First, combustion in RDEs is continuous: once initiated, detonation waves propagate azimuthally around the annular channel without requiring reignition, eliminating the need for repeated ignition. By contrast, pulse detonation engines (PDEs) operate in discrete cycles and must be reignited before each detonation, requiring a repeated ignition and purge sequence. Second, RDEs can operate across a wide range of flight conditions [10]. Whereas stable axial detonation propagation in an air-breathing engine generally requires a sufficiently high inflow velocity, azimuthal propagation allows RDEs to operate over a broad range of flight speeds.

Recent studies have investigated critical phenomena influencing combustion stability and performance in RDEs using experimental and numerical methods [11-13]. Smith and Stanley conducted experimental investigations on RDEs fueled with methane-oxygen, ethane-oxygen, and ethylene-oxygen mixtures [11] and examined the effects of equivalence ratio, mass flow rate, and nozzle geometry on thrust and specific impulse; converging nozzles improved performance by up to 13%. Prakash et al. performed high-fidelity numerical simulations on methane-oxygen RDEs [12] and showed that non-uniform reactant mixing significantly alters detonation strength and wave speed, while parasitic combustion downstream of the primary detonation front substantially degrades overall efficiency. Rankin et al. employed mid-infrared imaging to investigate hydrogen-air RDEs [13] and visualized detailed detonation structures such as reflected shock waves and shear layers while characterizing the mixing between combustion products and fresh reactants.

Complex flow and combustion phenomena in aerospace propulsion systems have been investigated using high-fidelity numerical simulations and advanced experimental diagnostics. Samie et al. examined three-dimensional coherent structures in a turbulent axisymmetric jet and demonstrated the roles of organized vortical structures and very-large-scale motions in shear-layer mixing and entrainment [14]. Malanoski et al. experimentally investigated a swirling lifted flame and showed that its unsteady response is governed by interactions among vortical disturbances, acoustic forcing, and flame-leading-edge motion [15]. Shin et al. analyzed DNS data for a lifted turbulent dimethyl ether (DME) jet flame and showed that fluid age, together with mixture fraction, provides a useful basis for characterizing reactive scalar distributions [16]. Among the computational tools used for such investigations, OpenFOAM provides an open-source and extensible framework for studying a wide range of aerospace flow and combustion problems.

OpenFOAM is widely used to solve fluid-flow problems [17]. Owing to its versatility, it has been applied to a broad range of aerospace flow problems [18-20]. Shin and Kim developed an enhanced pressure-based coupled algorithm within OpenFOAM by combining pressure-velocity and pressure-enthalpy coupling with the Kurganov-Tadmor flux-splitting scheme, enabling accurate simulations across a broad range of Mach numbers, from subsonic to supersonic regimes [18]. You et al. developed an automated aerodynamic analysis system for missile configurations by integrating OpenFOAM with open-source geometry modeling and mesh generation tools, demonstrating its practical applicability to industrial-level aerospace design workflows [19]. Li et al. employed OpenFOAM-based large-eddy simulation to investigate the aerodynamic effects of suction and ejection slits on airfoils at low Reynolds numbers, and demonstrated improvements in the lift-to-drag ratio through flow control [20].

Various OpenFOAM solvers have been developed specifically for detonation simulation [21-23]. Marcantoni et al. developed *rhoCentralRfFoam* and investigated the feasibility of simulating detonation cellular structures [21, 22]. The simulations reproduced the cellular structures and showed good agreement with those reported by Kirillov et al. [24]. The aspect ratios of the cellular structures computed on the extended domain agreed with the experimental measurements from Lefebvre et al. [25]. They also suggested that the cellular structure becomes independent of the initial perturbations as the detonation propagates. Weng et al. developed *RSDFoam*, a solver for real-gas shock and detonation simulations, by integrating *blastFoam* and Cantera [23, 26, 27]. The Redlich-Kwong (RK) equation of state (EoS) and the Noble-Abel (NA) EoS were implemented to account for real-gas behavior at high pressures, resulting in better agreement with reference data in the validation cases.

RDEs have also been studied with OpenFOAM-based solvers [28-30]. Xia et al. investigated mode transitions in RDEs using *detoFoam*, focusing on the shift from a single detonation wave with counter-rotating shock waves to two detonation waves [28]. They found that counter-rotating shock waves can ignite an additional detonation wave when they gain more energy—through interactions with the primary detonation wave and fresh reactants—than they dissipate elsewhere. Under these conditions, they intensify sufficiently to initiate the additional wave. Sheng et al. investigated the impact of multiple detonation waves on the stability and performance of RDEs using *BYCFoam* [29]. As the number of detonation waves increases, RDEs become more stable and achieve improved flow uniformity at the expense of reduced pressure gain at the outlet; thus, a balance between uniformity and pressure gain is required. Chen et al. investigated how equivalence ratio affects RDE propagation and performance using *rhoReactingCentralFoam* with a discrete inlet [30, 31]. The detonation mode shifts from a single-wave to a multi-wave mode as the equivalence ratio increases, with fuel-rich conditions enabling more rapid establishment of a stable detonation while altering thrust performance. Their findings emphasize that optimal RDE performance depends on balancing detonation stability with propulsion metrics such as specific thrust and specific impulse.

Despite these advances, most existing OpenFOAM-based detonation and RDE solvers employ density-based formulations. Pressure-based compressible formulations have nevertheless demonstrated applicability over a broad range of Mach numbers in other flow problems. Shin and Kim developed an enhanced pressure-based coupled algorithm within OpenFOAM by combining pressure–velocity and pressure–enthalpy coupling with the Kurganov–Tadmor flux-splitting scheme, demonstrating its applicability to flows ranging from subsonic to supersonic regimes [18]. Muralha et al. verified the pressure-based compressible flow solver ReFRESCO using multiple aerodynamic benchmark cases and reported results consistent with those obtained using the density-based solvers CFL3D and FUN3D [32]. However, the applicability of OpenFOAM's built-in pressure-based *multicomponentFluid* solver to shock-dominated reacting flows, particularly detonation and RDE flows, has not yet been systematically assessed.

In assessing this applicability, the limitations of the numerical methods adopted in the present framework must be considered. The solver does not employ a dedicated shock-capturing scheme and instead relies on highly diffusive numerical settings, including first-order upwind discretization for momentum convection and first-order implicit Euler time integration. Consequently, resolving the detailed internal structure of detonation waves requires very fine grids, imposing a substantial computational burden on RDE simulations. For future applications to integrated RDE–turbine simulations, system-level performance assessment does not necessarily require fully resolving these detailed structures, but it does require reliable prediction of the overall RDE flow characteristics and global performance quantities. The present study therefore aims to assess whether the pressure-based solver can provide such predictions at computationally practical grid resolutions, as a first step toward such integrated simulations.

Adaptive mesh refinement (AMR) and dynamic load balancing (DLB) are important for reducing the computational cost of RDE simulations. AMR dynamically refines the mesh during a simulation according to a prescribed criterion, such as the magnitude of the density gradient. Even when the assessment focuses on global flow and performance quantities, sufficient local resolution is needed around the detonation front and the interface between fresh reactants and burned products, where deflagration may occur. Because the regions requiring fine grids occupy only a small fraction of the RDE domain, AMR can apply high resolution selectively while retaining coarser grids elsewhere. DLB is essential in parallel AMR simulations because refinement concentrates cells on particular processors. Moreover, processors assigned to reacting regions incur additional computational cost from chemical-kinetics integration. These effects can create computational bottlenecks. DLB mitigates such imbalances by redistributing the computational domain at prescribed intervals or when the maximum load imbalance exceeds a specified threshold, thereby reducing the overall simulation time.

Several OpenFOAM-based solvers incorporate both AMR and DLB [33, 34]. Sun et al. developed *detonationFoam*, a solver for multi-species reactive flows with detailed finite-rate chemistry and an HLLC-P shock-capturing scheme [33]. The solver integrates AMR and DLB to improve computational efficiency. In oblique detonation wave simulations, AMR reduces the number of mesh cells by a factor of up to six, and the combination of AMR and DLB reduces CPU time by a factor of up to five while maintaining accuracy. Chen et al. developed *CDSFoam*, an Eulerian-Lagrangian solver for gas-droplet two-phase detonation combustion that employs an improved HLLC-LM approximate Riemann solver and detailed chemistry [34]. With AMR and DLB, *CDSFoam* significantly reduces computational cost while accurately predicting key detonation parameters and droplet dynamics. Beyond OpenFOAM-based solvers, Henry de Frahan et al. developed PeleC, an open-source compressible reacting-flow solver based on the AMReX framework [35]. PeleC employs block-structured AMR and supports scalable CPU- and GPU-based computations of large-scale reacting flows. Table 1 summarizes the OpenFOAM-based detonation solvers discussed above, together with additional solvers, highlighting their numerical schemes, base solver, and key features.

AMR and DLB will become increasingly important as engine development requires higher-resolution RDE simulations. However, the built-in dynamic mesh-refinement capability in OpenFOAM v12 does not directly support strictly two-dimensional meshes. Consequently, applying AMR to a nominally two-dimensional RDE problem requires artificial extrusion in the spanwise direction, which introduces unnecessary cells and additional computational cost. This requirement also complicates direct comparisons between uniform and adaptively refined two-dimensional simulations. To overcome this limitation, a two-dimensional AMR capability is implemented in the present study, enabling dynamic refinement directly in planar RDE domains. The developed two-dimensional AMR capability is then combined with the built-in DLB functionality of OpenFOAM v12 to improve parallel computational efficiency.

The objective of this study is to establish and systematically validate an open-source pressure-based framework for RDE simulations using OpenFOAM v12 and to evaluate the accuracy and computational efficiency of the developed two-dimensional AMR capability when combined with the built-in DLB functionality. The framework implements an RDE inlet boundary condition and a two-dimensional AMR capability while employing the built-in *multicomponentFluid* solver and DLB functionality of OpenFOAM v12. The framework is validated using simulations of one-dimensional planar detonation, two-dimensional cellular detonation, and a two-dimensional RDE. Its predictive capability is quantitatively assessed in terms of detonation velocity, mass flux, specific thrust, and specific impulse. Finally, uniform, AMR, and AMR+DLB simulations are compared to examine solution consistency, cell-count reduction, parallel load distribution, and computational speedup.

Table 1. Summary of features of various solvers for detonation simulation.

| Solver | Solver type | Base solver | Numerical schemes | | Reference |
|---|---|---|---|---|---|
| | | | Spatial discretization | Time discretization | |
| *rhoHLLCFoam* | Density-based | Unspecified | Harten-Lax-van-Leer-Contact (HLLC) | Crank-Nicholson | Liu et al. [36] |
| *ddtFoam* | Density-based | Unspecified | HLLC | Unspecified | Ettner et al. [37] |
| *blastFoam* | Density-based | Unspecified | HLL, HLLC, AUSM+, Tadmor/Kurganov | Euler, RK2, RK2-SSP, RK3-SSP, RK4, RK4-SSP | Heylmun et al. [38] |
| *detoFoam* | Density-based | Unspecified | Kurganov, Noelle and Petrova (KNP) | Unspecified | Xia et al. [28] |
| *BYCFoam* | Density-based | *rhoCentralFoam* | HLL, HLLC, AUSM+M, Kurganov | Euler | Sheng et al. [29] |
| *detonationFoam* | Density-based | *rhoCentralFoam* | HLLC-P | Euler | Sun et al. [33] |
| *RSDFoam* | Density-based | *blastFoam* | HLL, HLLC, AUSM+, Tadmor/Kurganov | Euler, RK2, RK2-SSP, RK3-SSP, RK4, RK4-SSP | Weng et al. [23] |
| *rhoCentralRfFoam* | Density-based | *rhoCentralFoam* | KNP | Euler | Marcantoni et al. [21] |
| *CDSFoam* | Density-based | *rhoCentralFoam* | HLLC-LM | RK3-SSP | Chen et al. [34] |
| *rhoReactingCentralFoam* | Density-based | *rhoCentralFoam* | KNP | Crank-Nicholson | McGough et al. [31] |
| *DCRFoam* | Density-based | Unspecified | AUSM+ -up | RK4 | Jiang et al. [39] |
| *UMdetFoam* | Density-based | Unspecified | HLLC | RK4 | Sato et al. [40] |

# 2. Numerical Methods

*2.1. Numerical schemes*

OpenFOAM v12 provides *multicomponentFluid* solver module for steady or transient turbulent flow of compressible multicomponent fluids with optional mesh motion and change [41]. The solver was used to compute RDE simulations in the present study. The detailed set of conservation equations of mass, momentum, energy, and species are provided in Appendix A.

OpenFOAM provides various numerical schemes for *multicomponentFluid* solver. Table 2 summarizes the features of the solver adopted in the present study. First, unlike the density-based solvers summarized in Table 1, the *multicomponentFluid* solver used here employs a pressure-based formulation, which sets it apart from the alternative approaches. For convection terms, the first-order upwind scheme is applied to the momentum equations, while the van Leer scheme is adopted to the remaining equations [42]. The first-order implicit Euler scheme is used for time discretization. Additionally, the second-order central differencing scheme is utilized for diffusion terms. These features are distinct from the features of *rhoHLLCFoam*, used for the validation simulation performed by Liu et al. [36] in Section 3.4 (see the first row in Table 1). The first-order upwind scheme was adopted for the momentum equations to ensure robust calculation of the strong pressure and velocity gradients associated with detonation-wave propagation. This choice may introduce additional numerical diffusion in slip lines and other fine-scale structures; therefore, the present numerical settings are primarily intended for evaluating global RDE flow and performance quantities.

Table 2. Summary of *multicomponentFluid* solver's features.

| Solver | Solver type | Numerical scheme | |
|---|---|---|---|
| | | Spatial discretization | Time discretization |
| *multicomponentFluid* | Pressure-based | Upwind, van Leer | Euler |

The discretized governing equations are solved using the PIMPLE algorithm provided by OpenFOAM [43]. The PIMPLE algorithm combines the features of both SIMPLE and PISO algorithms [44, 45]. Initially, the density correction is performed to prepare for subsequent computations. Using the corrected density, an intermediate velocity field is calculated in the momentum predictor step. Afterwards, the energy and species equations are solved. Following these steps, the PISO loop, constituting the inner loop of the PIMPLE algorithm, ensures mass conservation. Upon the completion of the inner loop, the algorithm returns to the pre-predictor step, initiating the next iteration within the outer loop. To promote sufficient pressure–velocity coupling at each time step, the PIMPLE algorithm was configured with three outer iterations and two inner pressure-correction iterations per outer iteration, using fixed iteration counts without residual-based early termination.

The detailed chemistry model is used to calculate chemical reactions. The San Diego mechanism is employed for reactions and thermo data [46]. Since hydrogen/air detonations are simulated in the present study, a simplified mechanism consisting of 24 reactions with 12 species is considered.

The reduced hydrogen mechanism and unity-Lewis-number assumption reduce the computational cost of the RDE simulations. These modeling assumptions may affect ignition delay, induction- and reaction-zone thicknesses, species diffusion, and the detailed transverse-wave and cellular structures. Therefore, the fine-scale detonation structures predicted in the present study may depend on the selected chemical mechanism and transport model. The primary objective of the RDE simulations is to evaluate global quantities. Assessment using more detailed chemical mechanisms and multicomponent transport models remains a subject for future work.

### *2.2. Inlet conditions of RDEs*

Inlet boundary properties ($p$, $T$, $\mathbf{u}$, $Y_i$) are determined by the difference between the bottom pressure, $p_{bot}$, the inlet stagnation pressure, $p_0$. We implemented the following calculation steps for the inlet boundary properties, provided by Schwer et al. [47], in OpenFOAM v12. Assuming isentropic expansion through convergent micro-nozzles at each inlet face, the inlet properties can be calculated from $p_{bot}$, $p_0$, and a critical pressure from choked condition, $p_{cr}$. The critical pressure, $p_{cr}$, is calculated as follows:

$$p_{cr} = p_0 \left( \frac{2}{\gamma + 1} \right)^{\frac{\gamma}{\gamma - 1}} \tag{1}$$

The calculation of the properties is divided into three conditions:

1. For $p_{bot} > p_0$, no injection occurs. *zeroGradient* boundary conditions are applied for $p$, $T$, and $Y_i$ and *noSlip* condition is applied for $\mathbf{u}$.
2. For $p_{cr} \le p_{bot} \le p_0$, the inlet properties are calculated as follows:

$$p = p_{bot} \tag{2}$$

$$T = T_0 \left( \frac{p_{bot}}{p_0} \right)^{\frac{\gamma-1}{\gamma}} \tag{3}$$

$$v = \sqrt{\frac{2\gamma}{\gamma-1} R T_0 \left[ 1 - \left( \frac{p_{bot}}{p_0} \right)^{\frac{\gamma-1}{\gamma}} \right]} \tag{4}$$

Stoichiometric $H_2$/air is applied to mass fraction boundary conditions.

3. For $p_{bot} < p_{cr}$, the flow is choked. The properties are fixed at critical values.

$$p = p_{cr} \tag{5}$$

$$T = T_{cr} = T_0 \left( \frac{p_{cr}}{p_0} \right)^{\frac{\gamma-1}{\gamma}} \tag{6}$$

$$v = v_{cr} = \sqrt{\frac{2\gamma}{\gamma-1} R T_0 \left[ 1 - \left( \frac{p_{cr}}{p_0} \right)^{\frac{\gamma-1}{\gamma}} \right]} \tag{7}$$

Also, stoichiometric $H_2$/air is applied to mass fraction boundary conditions.

### *2.3. Development of a two-dimensional adaptive mesh refinement method*

To enable adaptive refinement on a planar mesh with a single cell in the spanwise direction, a two-dimensional AMR method was developed by extending the topology-change framework of OpenFOAM v12. The standard AMR implementation employs the hexRef8 algorithm, which subdivides a hexahedral cell in all three coordinate directions to generate eight child cells. In the present implementation, this procedure was replaced with a planar four-way subdivision algorithm, denoted as hexRef4, while the overall mesh-refinement framework of OpenFOAM was retained.

The planar subdivision performed by the hexRef4 algorithm is illustrated in Figure 1, where the newly added points and edges are indicated by red dots and blue line segments, respectively. Only the edges lying in the x–y plane are bisected, whereas the edges aligned with the spanwise direction remain unchanged. The front and back faces are each divided into four coplanar faces, while each lateral face is divided into two, thereby subdividing one parent cell into four child cells without introducing a cell-center point. Consequently, refinement is confined to the computational plane, preserving both the single-cell layer in the spanwise direction and the empty boundary condition on the front and back boundaries.

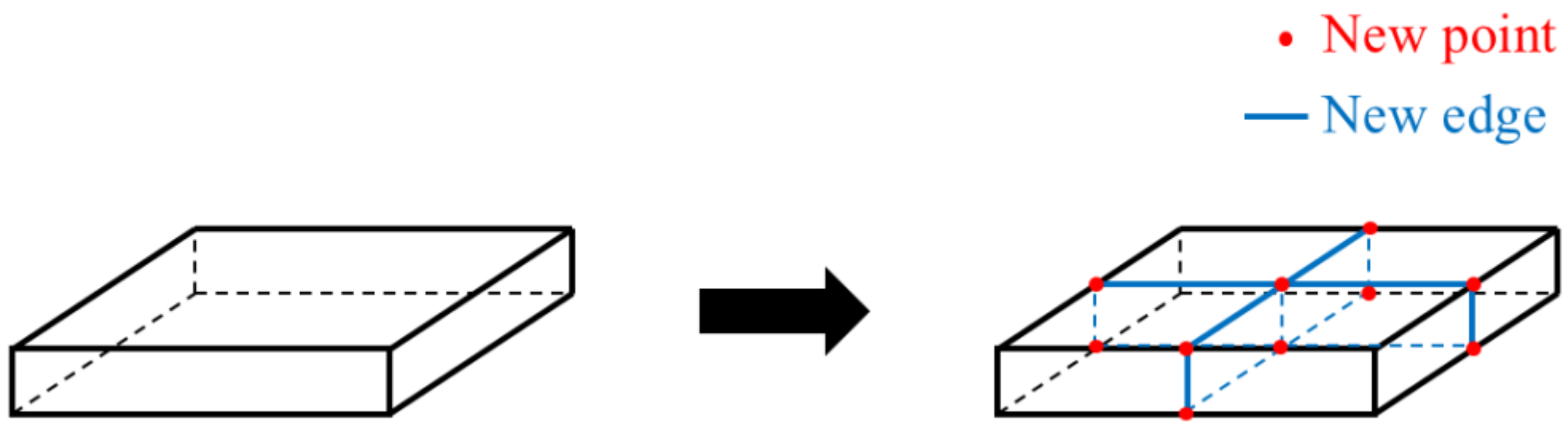

Figure 1. Schematic of the planar four-way subdivision of a hexahedral cell performed by the hexRef4 algorithm. Red dots and blue lines denote newly added points and edges, respectively.

The corresponding unrefinement procedure was also modified for the four-cell topology, such that four sibling cells originating from the same parent cell are merged into a single cell. The cell- and point-level consistency procedures of the original OpenFOAM framework were retained, including the constraint on the refinement-level difference between neighboring cells. During both refinement and unrefinement, the existing topology-mapping framework transfers the solution fields between the original and modified meshes. This mapping procedure was extended to exclude the zero-sized boundary fields associated with the front and back empty patches, allowing topology changes to be performed consistently on a two-dimensional mesh.

To support parallel computation, the original refinement-history framework was extended to refinementHistory2D, which records the parent–child relationship associated with the four-cell subdivision. The corresponding hexRef4Data structure preserves the cell and point refinement levels and the refinement history during mesh decomposition, reconstruction, and redistribution. The parallel decomposition and reconstruction procedures were therefore modified to read, transfer, and write these two-dimensional refinement data, ensuring that the mesh hierarchy remains consistent when refined cells are transferred between processor domains. This extension enables the developed two-dimensional AMR method to operate with the existing dynamic load-balancing functionality of OpenFOAM without modifying the load-balancing algorithm itself.

# 3. Results and Discussion

In this section, we present validation and performance results obtained from a sequence of increasingly complex test cases. We begin with a one-dimensional detonation tube configuration, for which a detailed grid-convergence study is performed and the simulation outcomes are validated against the experimental measurements of Malik et al. [48]. Next, we assess the capability of the framework to capture detonation cell structures in two-dimensional simulations by comparing the predicted cell sizes and patterns with established results from Oran et al. [49]. Building on this, a two-dimensional rotating detonation engine (RDE) configuration is examined, and the computed performance metrics—including specific thrust, specific impulse, and mass flux—are compared with previous numerical results reported by Liu et al. [36], along with pressure distributions and detonation velocities. Finally, we then investigate the effectiveness of adaptive mesh refinement (AMR) and dynamic load balancing (DLB) in reducing computational cost, verifying consistency with the baseline RDE simulations and analyzing the distribution of computational cells across processors.

Additionally, to highlight the potential of the framework for capturing three-dimensional effects, we conduct unwrapped RDE simulations with spanwise equivalence-ratio variations and examine their influence on the detonation front in Appendix B, thereby providing a proof of concept toward full three-dimensional modeling.

### *3.1. Validation on one-dimensional planar detonation*

A one-dimensional planar detonation problem is selected to validate the solver. Malik et al. conducted experiments in a detonation tube with premixed $H_2/O_2/Ar$ in a wide range of equivalence ratio, measuring the pressure signal at 9 locations (0.5, 1.5, ..., 8.5 m), and the detonation velocity [48]. The simulation is performed under identical conditions along with a grid study.

#### *3.1.1. Problem setup*

Figure 2 illustrates the problem setup. A 9 m-long, one-dimensional domain is constructed for grid study and validation. The wall condition with no-slip is employed for the left boundary condition, whereas a *waveTransmissive* outlet boundary condition, minimizing the reflection of pressure waves, is adopted for the right boundary condition. Initially, the entire domain is filled with stoichiometric $H_2/O_2$ at $p_{initial} = 100$ kPa and $T_{initial} = 300$ K. A detonation wave is initiated by applying a high pressure and a high temperature ($p_{ignition} = 5$ MPa, $T_{ignition} = 3{,}000$ K) to the ignition region ($x \leq 0.01$ m) which is a similar ignition condition to Liu et al. [36]. Detonation is successfully formed and propagates in the simulation. As the current work aims to develop RDE simulation frameworks, the grid sizes are chosen based on Liu et al.'s RDE validation data [36]. The tested grid sizes are 0.2, 0.1, 0.05, 0.025, and 0.0125 mm.

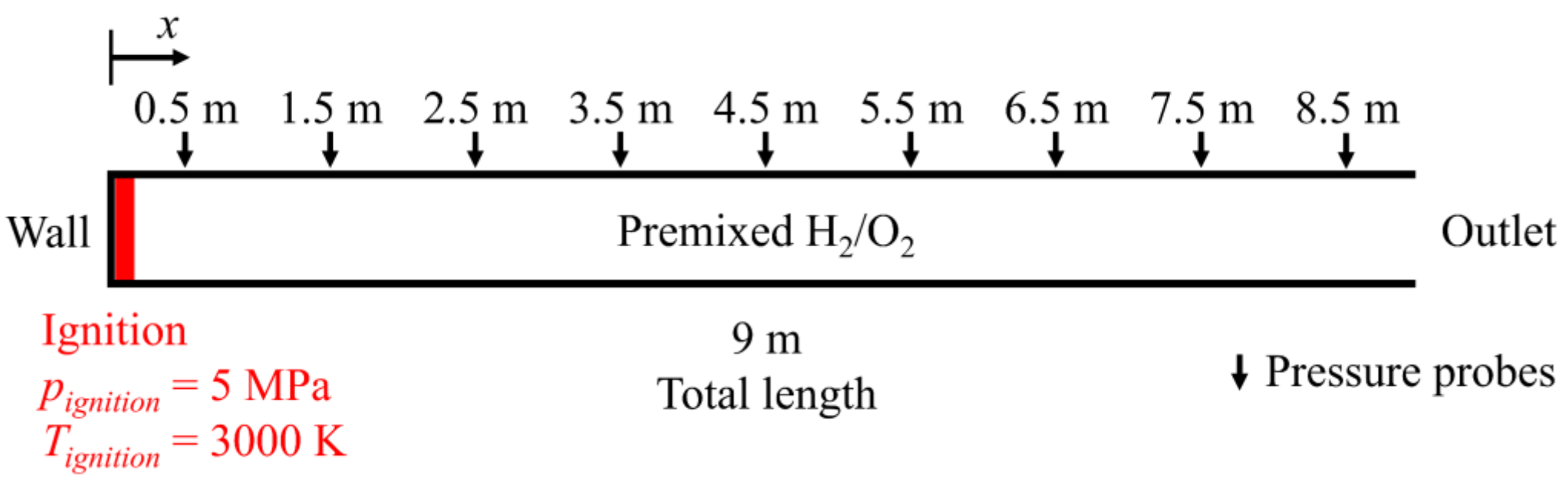

Figure 2. Computational domain for one-dimensional detonation tube.

*3.1.2. Simulation results*

Figure 3 presents snapshots of the temperature fields at times when the detonation wave reaches multiple locations of 0.5, 1.5, 4.5, and 7.5 m with the grid size of 0.1 mm. The detonation wave propagates steadily to the right. Upon passage, the temperature jumps are maintained sharp, while regions ahead of the wave front remain at the initial temperature, unaffected by the detonation.

Figure 4 shows the pressure distributions by different grids. On each grid, the pressure distribution is captured when the peak pressure is located at 4.5 m, magnifying differences in the detonation wave structure. On the right side of the detonation wave, the steepness of the leading shock increases as the grid size decreases. This occurs because the shock front is resolved by a relatively constant number of cells, regardless of grid resolution. For each case, the distance from the zero position to where pressure initially rises increases (from 0.05 mm to 0.8 mm), which is proportional to the respective grid sizes. On the left side, as the grid is refined, the pressure decay profiles converge. Simultaneously, the differences between simulations get smaller as the grid sizes of them decrease.

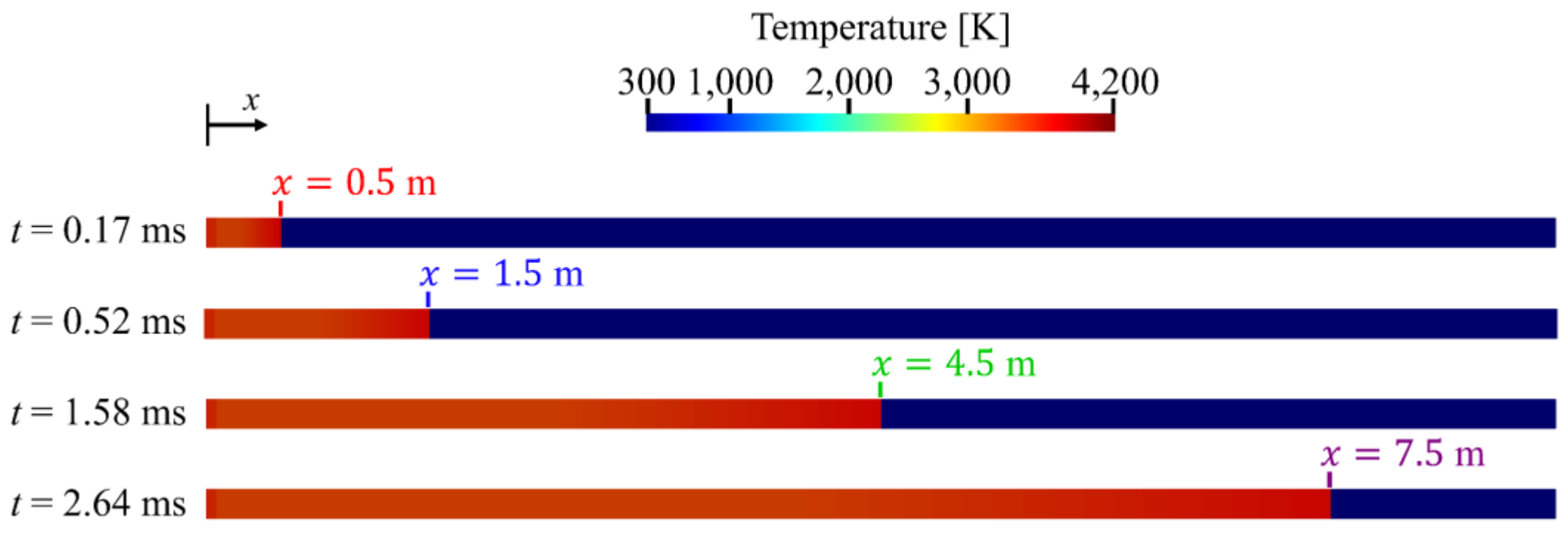

Figure 3. Snapshots of detonation propagation passing through the specified locations with $\Delta x = 0.1$ mm.

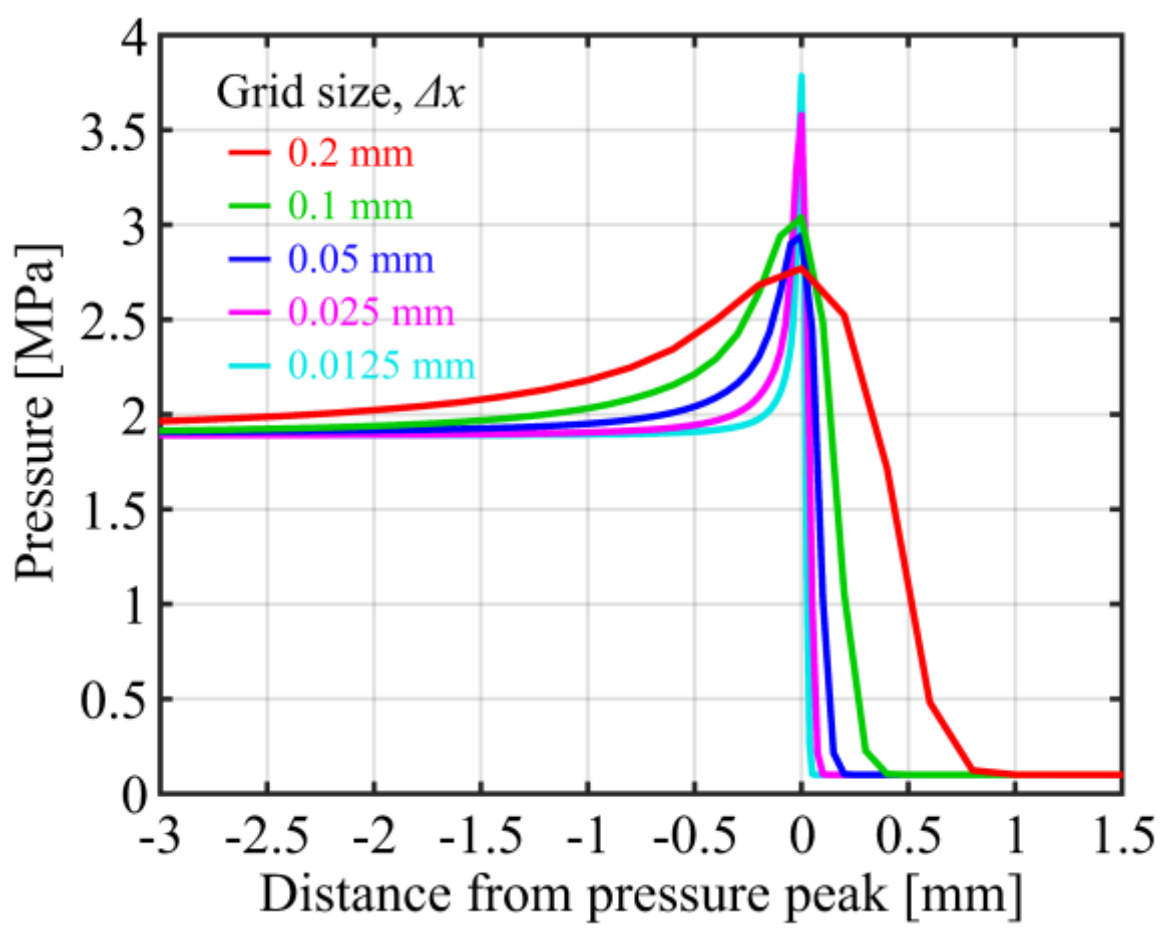

Figure 4. Magnified pressure profiles using different grid sizes when the detonation waves reach at $x$ = 4.5 m (see Figure 3).

Table 3 summarizes the average detonation velocities, which were obtained by fitting linear regressions to the detonation-wave position versus time. All grid resolutions yield similar detonation velocities around 2,832 m/s. Thus, we conclude that the base grid ($\Delta x = 0.1$) provides sufficient accuracy for further validation.

Table 3. Average detonation velocities by grid sizes and their relative errors to CJ velocity.

| | CJ velocity | Grid size, $\Delta x$ [mm] | | | | |
|---|---|---|---|---|---|---|
| | | 0.2 | 0.1 | 0.05 | 0.025 | 0.0125 |
| $D$ [m/s] | 2,834.96 | 2,832.79 | 2,832.22 | 2,832.43 | 2,832.19 | 2,832.00 |
| Relative error [%] | - | - 0.08 | - 0.10 | - 0.09 | - 0.10 | - 0.10 |

The simulation for the grid size of 0.1 mm is further validated with the experimental results of Malik et al.

[48]. Figure 5 illustrates the variation of pressure over time at the locations presented in Figure 3. Pressure variation is calculated as follows:

$$\Delta p = p - p_{atm}, \tag{8}$$

where $p$ is the simulated pressure, and $p_{atm}$ is the atmospheric pressure (100 kPa). The black lines represent experimental data, while the red, blue, green, and purple lines show the simulation results at $x$ = 0.5, 1.5, 4.5, and 7.5 m, respectively. To improve readability, the pressure curves at $x$ = 1.5, 4.5, and 7.5 m are vertically offset by 1, 2, and 3 MPa, respectively as well. The sudden increases in pressure as the detonation wave passes each location are well captured by the simulation. The subsequent pressure decay trends also show good agreement with the experimental data.

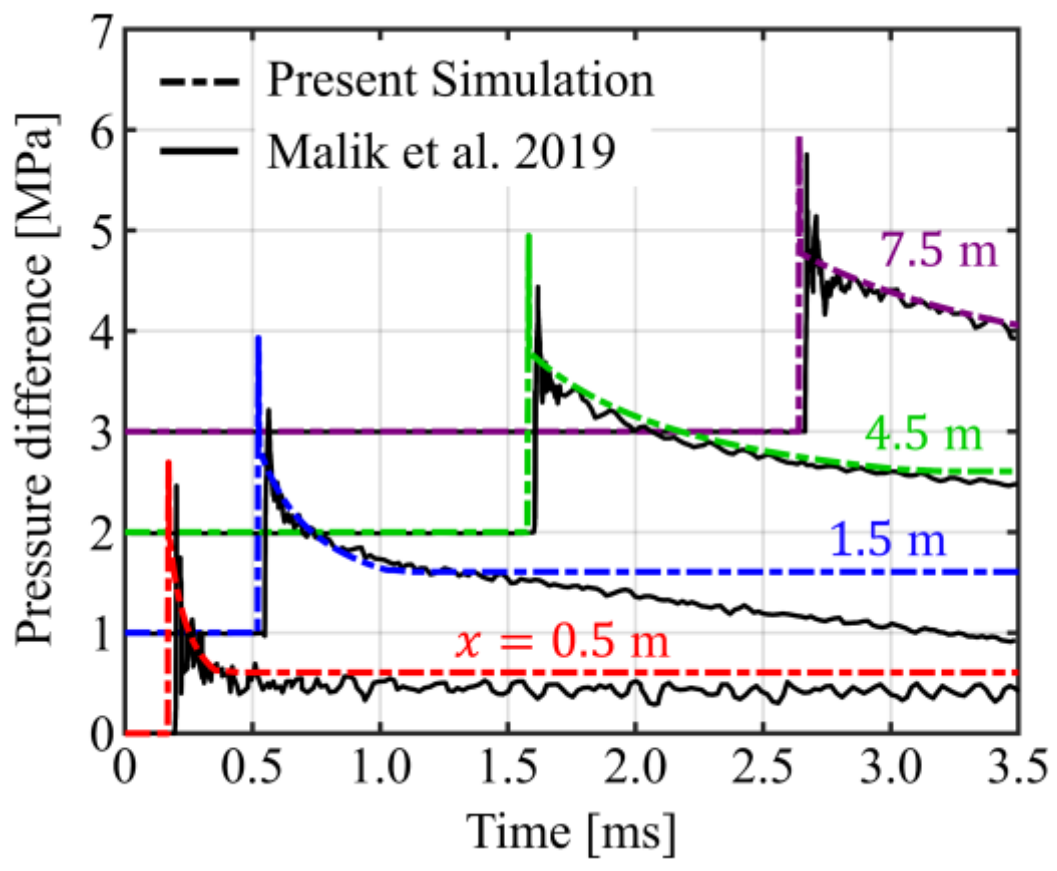

Figure 5. Comparison of pressure signals at the specified pressure probes.

Table 4 shows the comparison of the average detonation velocities and corresponding errors. The Chapman-Jouguet (CJ) detonation velocity was calculated by SD Toolbox [50], as well as the results of Malik et al. is compared [48]. In Table 4, the average detonation velocity of computed previously is also considered for validation. The present simulation reproduced the detonation velocity within 0.1 % error compared to Malik et al. and the CJ velocity. These results demonstrate that the solver accurately captures property variations during detonation combustion and the propagation of detonation waves.

Table 4. Comparison of average detonation velocity and the relative errors.

| Cases | $D_{avg}$ [m/s] | Relative error *w.r.t.* Malik et al. | Relative error *w.r.t.* CJ velocity |
|---|---|---|---|
| Malik et al. [48] | 2,830.49 | - | - |
| CJ velocity [50] | 2,834.96 | - | - |
| Present simulation with *Δx=0.1 mm* | 2,832.22 | 0.06 % | - 0.10 % |

### *3.2. Validation of the ZND induction-zone structure*

To examine whether the present solver can accurately reproduce the detailed ZND structure of a detonation wave, a one-dimensional simulation is compared with a steady ZND solution obtained using SDToolbox [50]. Particular attention is given to the thermodynamic profiles, species evolution, and induction-zone length, which are sensitive to numerical resolution and the treatment of finite-rate chemistry.

#### *3.2.1. Numerical setup*

To examine the detailed ZND structure, the grid spacing used in the one-dimensional detonation simulation of Section 3.1 was progressively reduced. Twelve uniform grid spacings of 4, 2, 1.75, 1.5, 1.25, 1, 0.5, 0.25, 0.125, 0.0625, 0.03125, and 0.015625 μm were considered. To reduce computational cost, the streamwise domain length was set to 20 mm for grid spacings from 4 to 0.125 μm and shortened to 6 mm for the three finest cases. The ZND profile obtained using SDToolBox was mapped onto the OpenFOAM mesh as the initial condition. A stoichiometric hydrogen-oxygen mixture diluted with argon ($H_2$:$O_2$:Ar = 2:1:7) was used at an initial temperature of 300 K and an initial pressure of 101.325 kPa. The remaining numerical methods and solver settings were the same as those used in Section 3.1.

#### *3.2.2. Simulation results*

Figure 6 compares the shock-aligned temperature, pressure, and density profiles with the corresponding SDToolbox ZND solution over the first 1 mm behind the leading shock. The black dashed lines denote the SDToolbox results. All grid resolutions exhibit overshoots in the thermodynamic variables immediately behind the leading shock. This common tendency appears to be associated with the limitation of the pressure-based formulation, in which the pressure field is obtained by solving a Poisson equation, when capturing the strong discontinuity at the detonation front. The coupled pressure, temperature, and density fields consequently exhibit similar near-shock deviations. After the overshoot, the temperature increases toward the equilibrium state, whereas the pressure and density decrease toward the corresponding ZND profiles. On the coarse meshes, the captured shock and the associated overshoot occupy a larger physical distance because the discontinuity is represented over a finite number of computational cells. The induction zone is therefore obscured and cannot be distinguished sufficiently from the leading-shock region. As the grid is refined, the physical extent of the captured shock and overshoot decreases, allowing the induction zone to be separated from the leading shock. The subsequent temperature rise and pressure and density relaxation progressively approach the ZND profiles. Apart from the captured discontinuity immediately behind the leading shock, the finest-grid results nearly overlap the ZND solution throughout the induction and principal reaction zones.

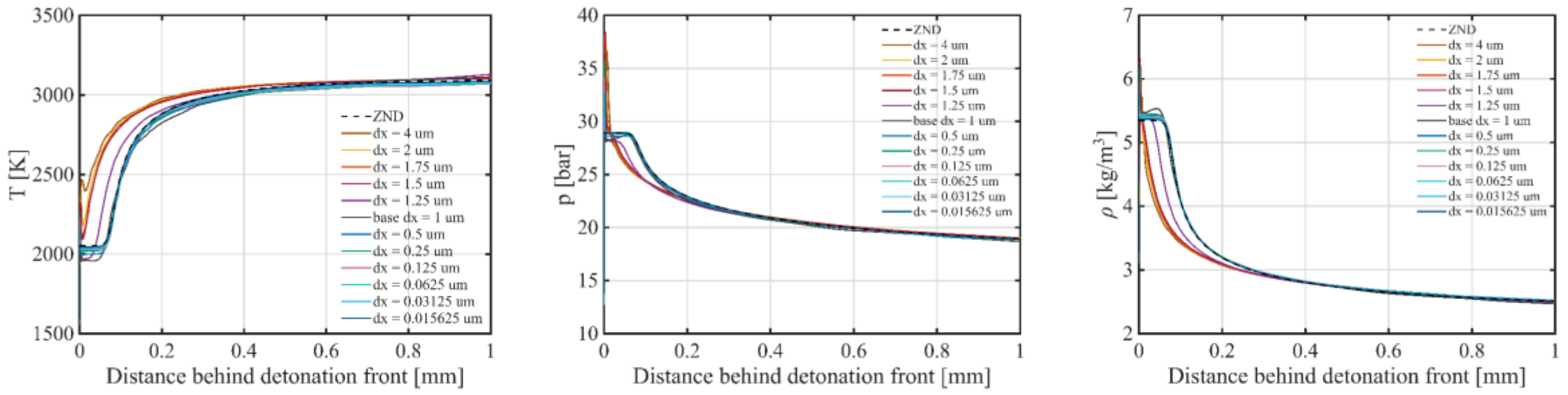

Figure 6. Shock-aligned profiles of temperature (left), pressure (center), and density (right) over the first 1 mm behind the leading shock. The black dashed lines denote the SDToolbox ZND solution.

For quantitative assessment, the induction length in each OpenFOAM simulation was evaluated as the distance from the leading shock to the location of maximum volumetric heat-release rate. This reaction-front identification follows the procedure adopted in previous detonation CFD grid-convergence studies [51]. The reference ZND induction length was calculated by SDToolBox.

As the grid was refined, the OpenFOAM induction length converged toward the ZND value. For grid spacings larger than 1 μm, the grid was too coarse to sufficiently capture the induction zone as the temperature and pressure decreased from their overshoots. A distinct induction region begins to emerge at a grid spacing of 1 μm, although its predicted length remains 8.90% below the ZND value. For grid spacings of 0.5 μm and smaller, the induction-length difference remains below 1%, indicating that the induction zone is quantitatively resolved. At the finest grid spacing of 0.015625 μm, the OpenFOAM induction length was 83.578 μm, compared with the ZND value of 83.425 μm. The difference was 0.154 μm, or 0.18%.

Despite the substantial grid dependence of the near-shock induction structure, all grid resolutions approach similar downstream thermodynamic states as the reaction proceeds toward the equilibrium CJ condition. This distinction is important when selecting a grid for multidimensional RDE simulations. Resolving the induction-zone structure requires a grid spacing of approximately 0.5 μm or smaller under the present conditions. In contrast, a coarser grid may remain adequate when the objective is limited to global RDE performance quantities, such as mean detonation velocity, chamber pressure, or integral performance, rather than the detailed internal structure of the detonation front.

Table 5. Grid-convergence results for the predicted induction-zone length compared with the ZND reference value.

| Grid size (μm) | Cells per $L_{ind,ZND}$ | $L_{ind}$ (μm) | Difference from ZND (%) |
|---|---|---|---|
| 4 | 20.9 | 8.000 | 90.41 |
| 2 | 41.7 | 20.000 | 76.03 |
| 1.75 | 47.7 | 22.749 | 72.73 |
| 1.5 | 55.6 | 24.001 | 71.23 |
| 1.25 | 66.7 | 53.750 | 35.57 |
| 1 | 83.4 | 76.000 | 8.90 |
| 0.5 | 166.8 | 83.000 | 0.51 |
| 0.25 | 333.7 | 83.000 | 0.51 |
| 0.125 | 667.4 | 83.250 | 0.21 |
| 0.0625 | 1,334.8 | 84.062 | 0.76 |
| 0.03125 | 2,669.6 | 83.750 | 0.39 |
| 0.015625 | 5,339.2 | 83.578 | 0.18 |
| ZND | - | 83.425 | - |

Figure 7 compares the mass-fraction profiles of H2, O2, H2O, and OH with the corresponding SDToolbox ZND solution. The coarse-grid results show noticeable deviations in the near-shock and reaction-zone profiles because the detailed ZND structure is not sufficiently resolved. As the grid is refined, the locations and shapes of the species profiles progressively approach the ZND solution. The finest-grid results closely follow the reference profiles throughout the induction and reaction zones, supporting the ability of the present solver to reproduce the species evolution within the ZND structure.

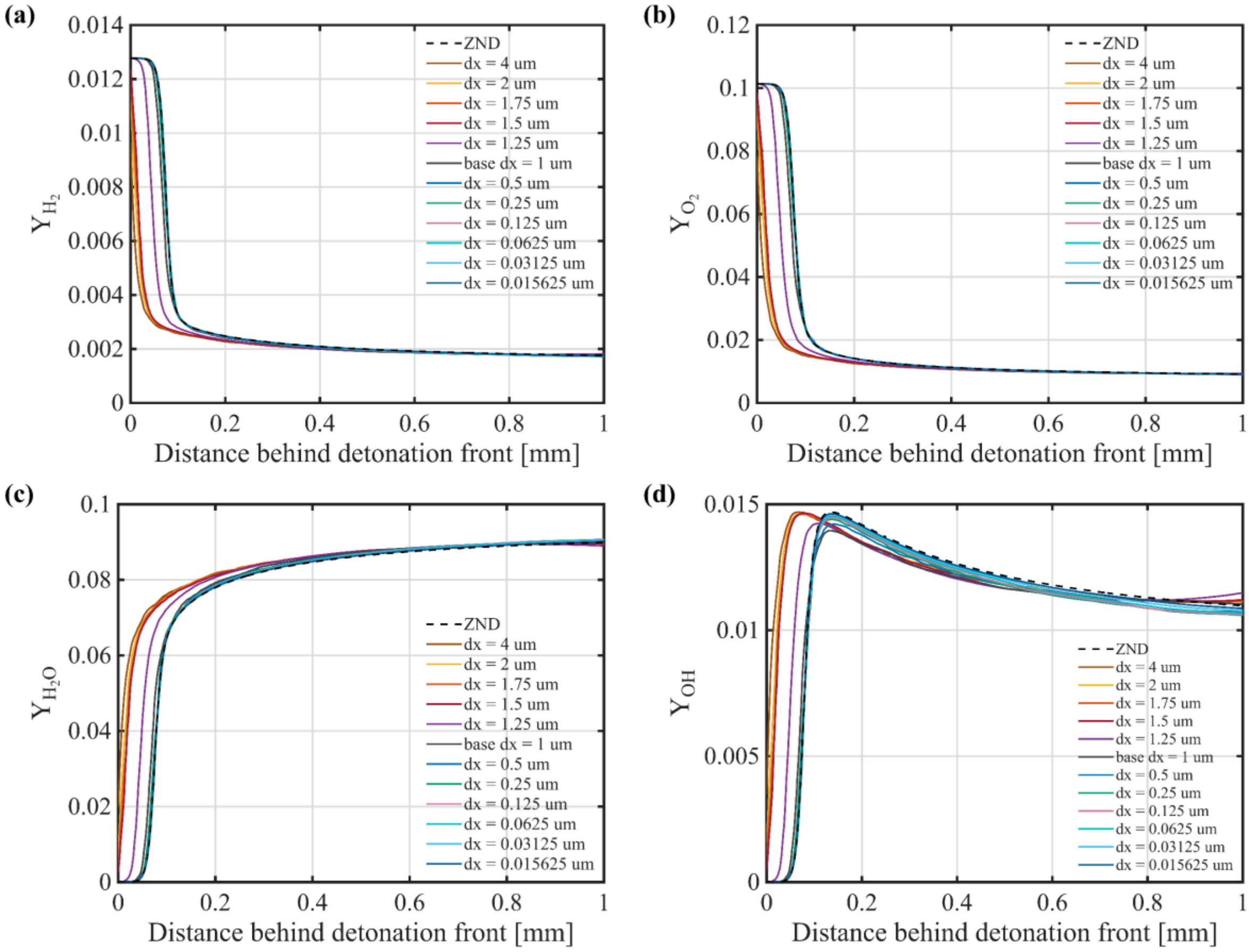

Figure 7. Shock-aligned mass-fraction profiles of (a) H2, (b) O2, (c) H2O, and (d) OH over the first 1 mm behind the leading shock. The black dashed lines denote the SDToolbox ZND solution.

### *3.3. Two-dimensional cellular structure of planar detonation propagation*

To establish the reliability of the present detonation simulations, the computed cell structure is compared with well-known results from the literature, particularly those reported by Oran et al. [49]. Validation against cell size and pattern is essential, since the interactions of transverse waves and triple points govern the global detonation dynamics, and an inaccurate reproduction of these features would undermine discussions of instability modes, wave curvature, or boundary effects. The present study evaluates the simulated cell structures and systematically compares them with the results reported by Oran et al. [49].

#### *3.3.1. Numerical setup*

The numerical setup for the two-dimensional planar detonation simulation is shown in Figure 8. To reproduce the detonation cell structure reported by Oran et al. [49], the initial temperature, pressure, and fuel–oxidizer composition were set to 298 K, 6,670 Pa, and $H_2$:$O_2$:Ar = 2:1:7, respectively. A one-dimensional CJ detonation solution was mapped onto the computational domain as the initial condition, and unreacted gas pockets were embedded within the post-detonation region to introduce the initial perturbation, as shown in Figure 8. A base grid spacing of 200 μm was used, and three levels of two-dimensional AMR were applied, resulting in a finest grid spacing of 25 μm. Mesh refinement was controlled using the density-gradient magnitude to resolve the detonation front and the associated transverse-wave structures.

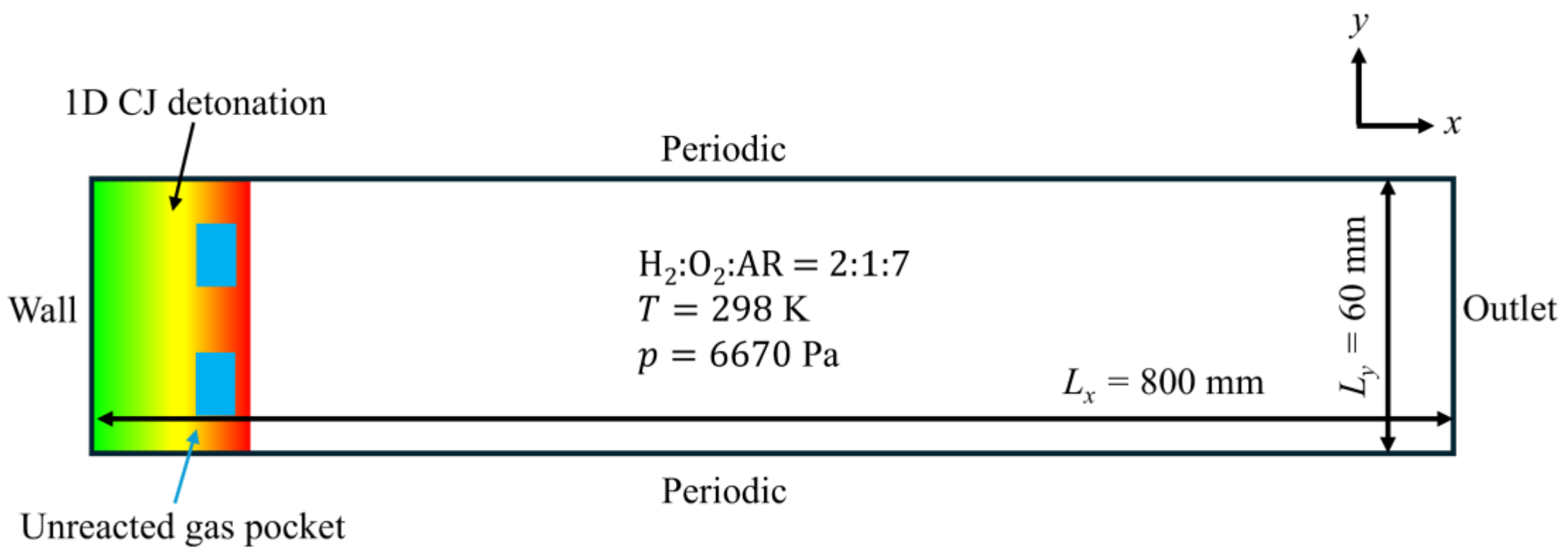

Figure 8. Computational domain for planar detonation cell simulations.

*3.3.2. Results and validation*

The instantaneous temperature and pressure fields in Figure 9 show the perturbed leading front; however, the triple-point structure is not clearly distinguishable in these snapshots. This is likely associated with the finite grid resolution and the limitations of the pressure-based formulation at the leading shock. As demonstrated in the ZND-structure validation in Section 3.2, the pressure-based solver produces an overshoot immediately behind the shock, and this behavior becomes more influential when the detonation structure is represented on a relatively coarse grid. The limited visibility of the triple points in the instantaneous fields does not imply the absence of a cellular structure, which is instead revealed by the time-integrated numerical soot foil in Figure 10.

A numerical soot foil was constructed from the maximum pressure recorded at each spatial location over the simulation period to visualize the triple-point trajectories. The vertical black lines in Figure 10 primarily reflect the pressure overshoot generated as the numerically captured leading shock passes through successive spatial locations. They should therefore not be interpreted as physical triple-point tracks. This numerical signature is consistent with the leading-shock overshoot observed in the ZND-structure results and represents a limitation of the pressure-based solver at the present grid resolution. Apart from these vertical numerical traces, the intersecting high-pressure tracks delineate the triple-point trajectories and clearly form the cellular structure shown in Figure 10.

The cellular structure obtained from Figure 10 does not reproduce the pattern reported by Oran et al. [49] exactly. Nevertheless, the numerical soot foil gives a representative detonation cell size of approximately 62 mm by 30 mm, compared with 54 mm by 31 mm reported by Oran et al. [49]. The transverse cell dimension is therefore reproduced closely, whereas a larger difference remains in the streamwise dimension. Considering the pressure overshoot and the limited resolution of the detailed reaction zone, the present comparison indicates that the solver captures the principal scale and propagation pattern of the cellular detonation, while fine-scale quantitative agreement remains sensitive to the shock treatment and grid resolution.

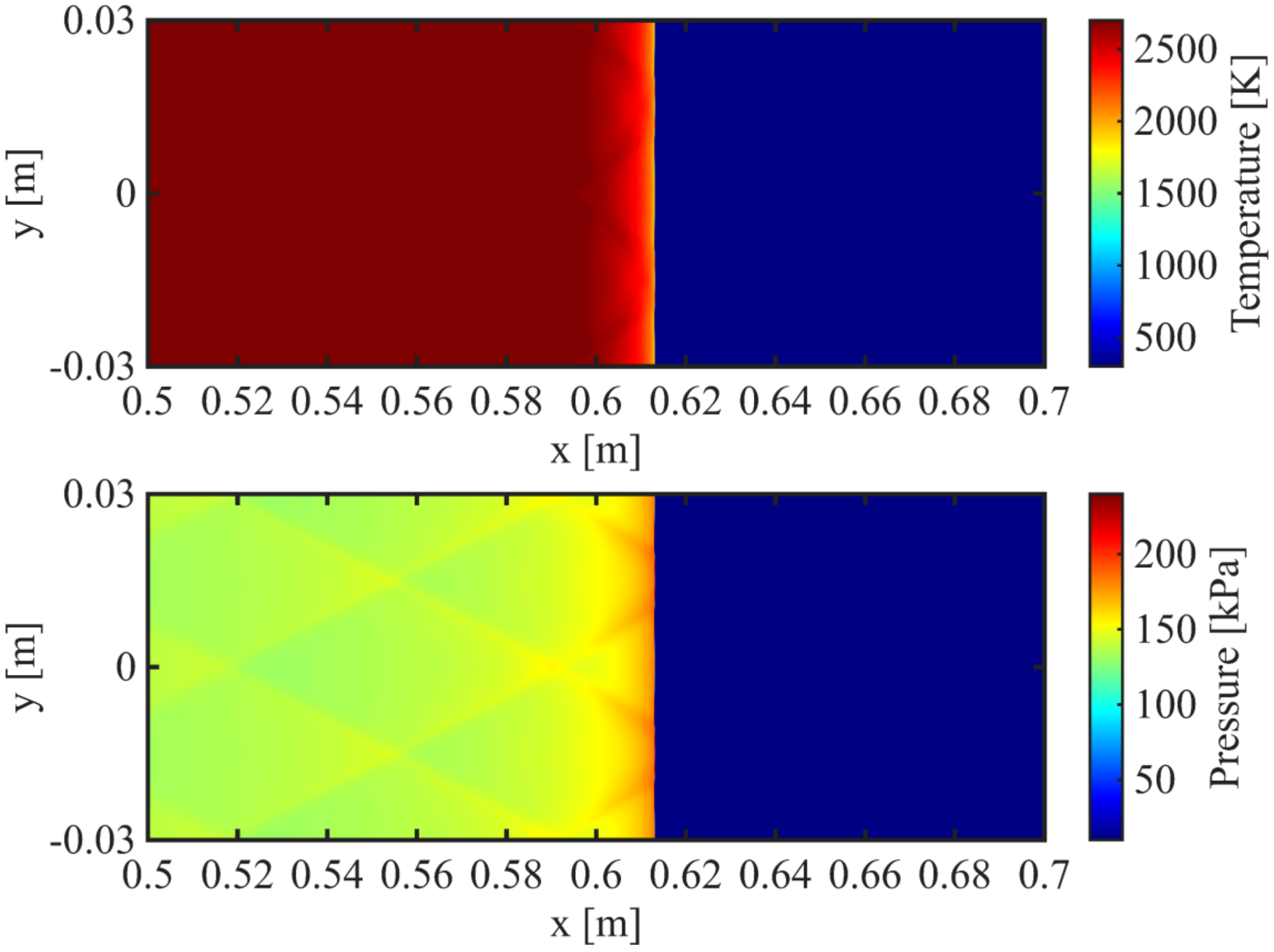

Figure 9. Instantaneous temperature (top) and pressure (bottom) fields at $t$ = 325 μs.

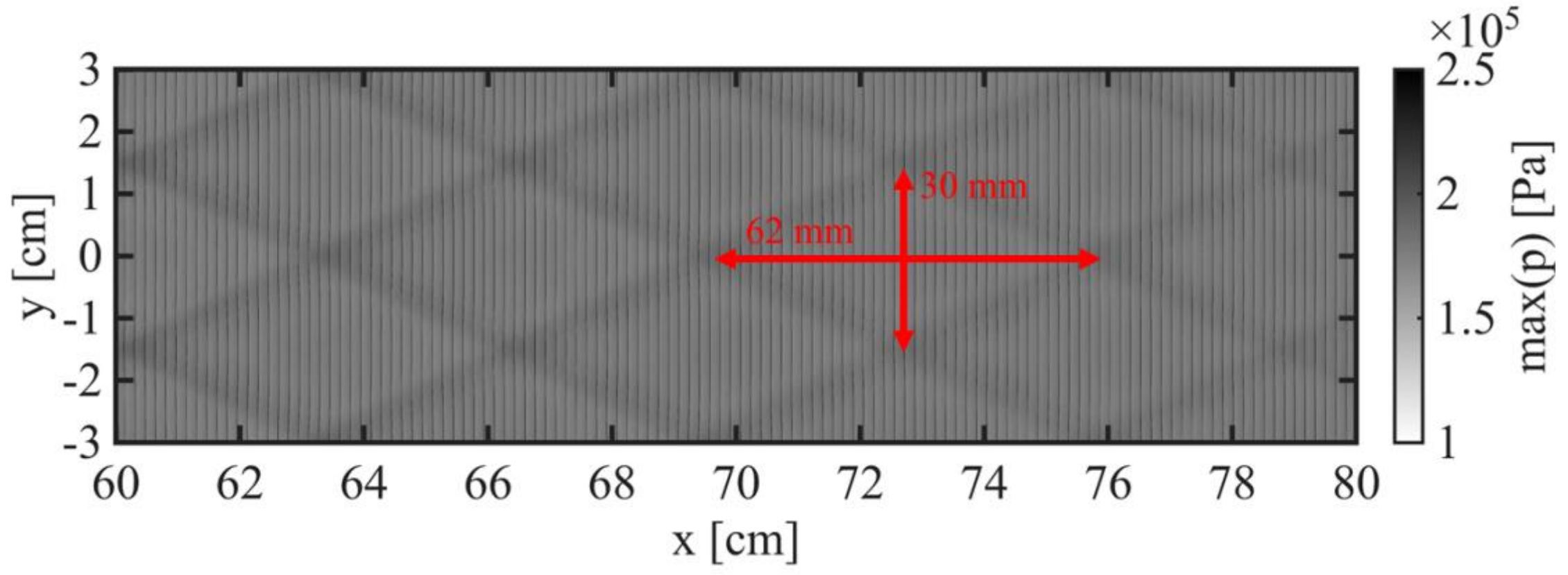

Figure 10. Numerical soot-foil pattern constructed from the maximum pressure recorded at each spatial location over the simulation period.

Table 6. Comparison of the detonation cell size between the present simulation and Oran et al. [49].

| | Cell size |
|---|---|
| Present simulation | 62 mm by 30 mm |
| Oran et al. [49] | 54 mm by 31 mm |

### 3.4. Two-dimensional RDE simulation

In this section, we develop a framework for two-dimensional RDE simulations. Due to the narrow annular channel geometry, the RDE flow field can be approximated as a two-dimensional plane with periodic boundary conditions applied on the lateral boundaries (see Figure 11). The simulation results are validated against another two-dimensional RDE simulation by Liu et al.[36].

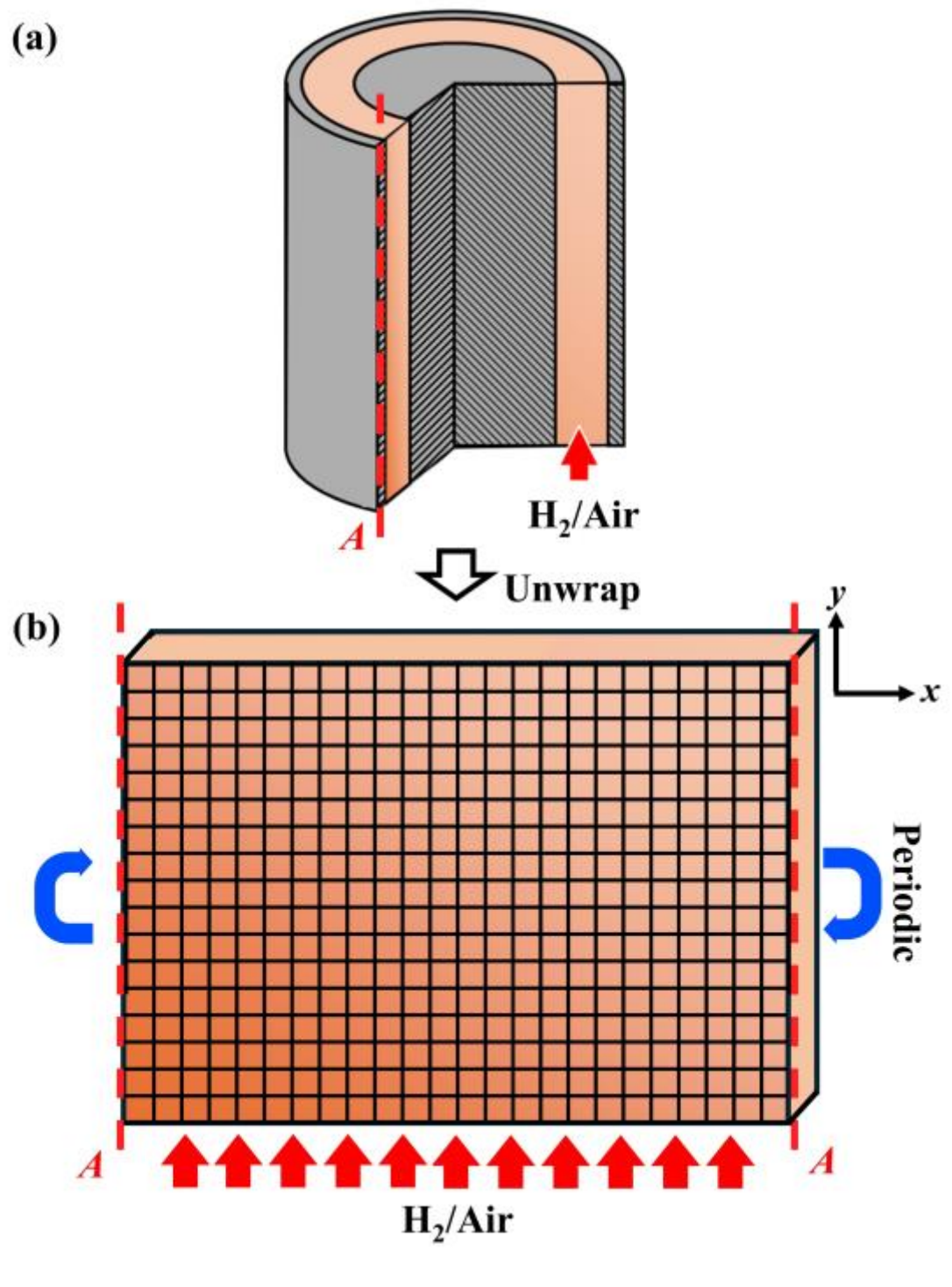

Figure 11. (a) An annular RDE configuration, and (b) a two-dimensional domain for unwrapped RDEs.

#### 3.4.1. Numerical setup

Figure 12 illustrates the numerical setup used in this study. The present simulations are conducted using a modified geometry from Liu et al. [36]. The original computational domain by Liu et al. consisted of an RDE with a 10 mm radius (hence a span of 62.83 mm) and a 30 mm height, initialized with two detonation waves. To reduce computational cost, we consider a half domain in the $x$-direction, covering a single detonation wave while preserving the RDE's area-averaged properties [36]. Additionally, the domain is extended in $y$ direction to minimize the effect by the outlet boundary conditions. The resulting domain measures 31.25 mm in the $x$ direction and 60 mm in the $y$ direction, with a uniform grid spacing of 0.0625 mm. At this resolution, the present RDE simulations focus on global flow and performance quantities and do not attempt to resolve the detailed ZND structure examined in Section 3.2.

For the inlet, a uniform inflow without an injector geometry (inlet-area ratio of 1) is used. The inlet pressure, temperature, and velocity are set by the formulas from Section 2.2. The reference values of the inlet stagnation temperature ($T_0$) and pressure ($p_0$) are 360 K and 1.0 MPa, respectively. Periodic boundary conditions are applied to the lateral sides to consider the cylinder shape of the RDE. Finally, *waveTransmissive* condition is applied to the outlet, minimizing the reflection of waves. For reactants, a stoichiometric mixture of $H_2$/Air is used.

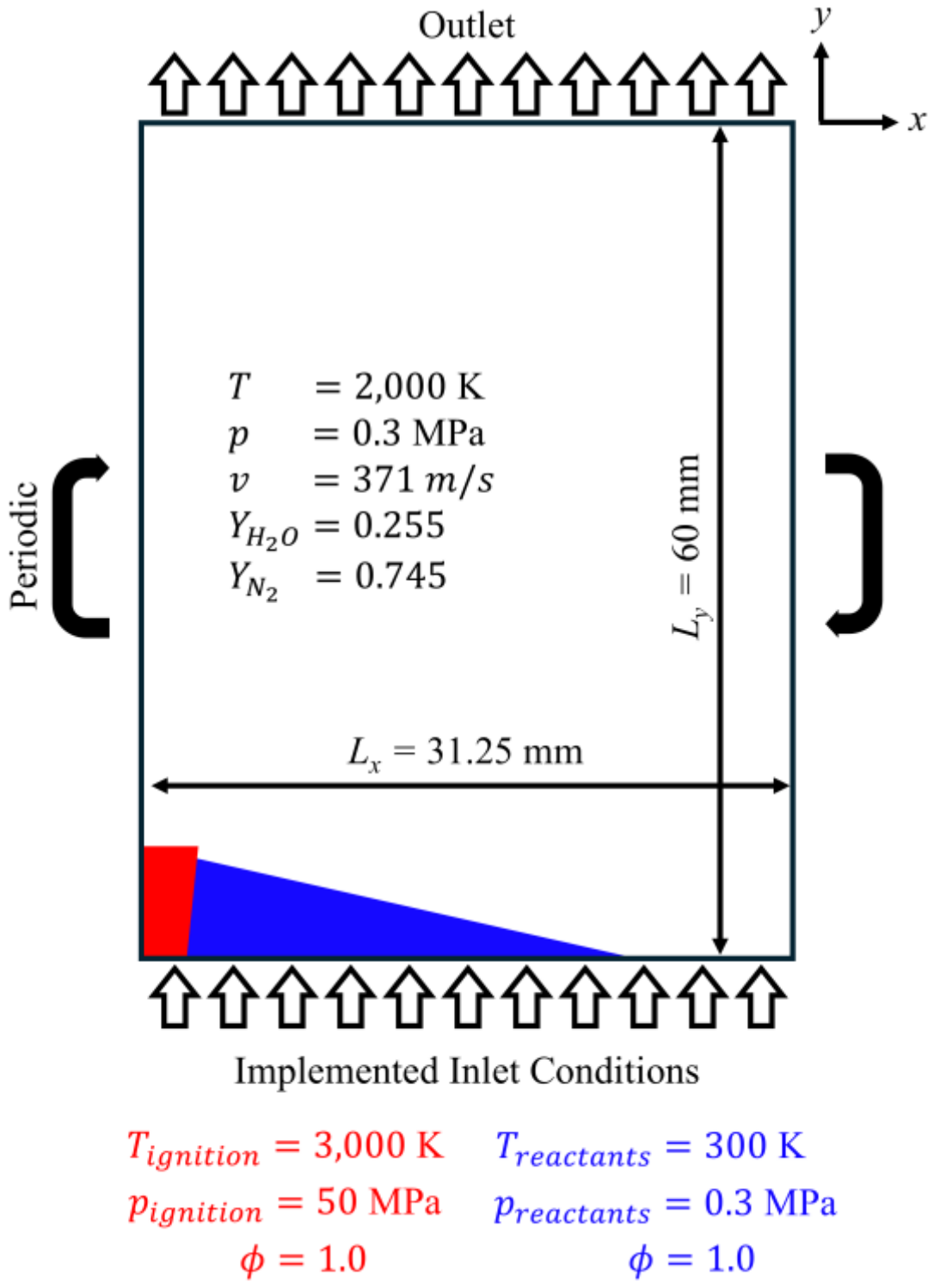

Figure 12. Configuration of numerical setup for two-dimensional RDE simulations.

The initial conditions are set by three regions as shown in Figure 12: the ignition zone in red, the reactants zone in blue, and the remainder in white of the domain. The ignition zone (red in Figure 12) is designated for direct initiation of detonation and is filled with elevated temperature and pressure of $T_{ignition} = 3{,}000$ K and $p_{ignition} = 50$ MPa and a premixed $H_2$/air mixture. The reactants zone (blue in Figure 12) is filled with a premixed $H_2$/air mixture at $T_{reactants} = 300$ K and $p_{reactants} = 0.3$ MPa for initial detonation propagation. Its height of the reactants zone tapers along the *x* direction to mimic the RDE's stable flow field and to promote fast convergence. The interface between the ignition zone and the reaction zone is slightly tilted to the right to align with the steady-state detonation wave front. Finally, the remaining region is filled with combustion products ($H_2O$ and $N_2$) at $T_{products} = 2{,}000$ K and $p_{products} = 0.3$ MPa. Regarding the velocity, a uniform velocity of $[0, 371]$ m/s, derived from Equation (7), is initially set on the entire domain. The simplified domain, uniform inlet, and initialization procedure were adopted consistently with Liu et al. [36] to enable a direct numerical-to-numerical comparison, while the extended axial domain was used to reduce the influence of the outlet boundary. The effects of these assumptions were not independently assessed and remain subjects for future sensitivity studies.

### *3.4.2. Results and validation*

Simulation is conducted for 500 μs and the detonation wave circles the whole domain 28 times. The region under $y = 30\ mm$ is considered and shown to validate the present simulation. Figure 13 illustrates exemplary flow fields of the simulation from 190 μs to 202 μs. The left graphs show temperatures on a linear scale, while the right graphs show pressures on a log scale at the same time steps. As time goes by, a single detonation wave propagates to the right illustrating typical RDE structures described below.

The typical characteristic features of RDEs are well captured as shown in Figure 13, labeling key regions by (a)-(f). At location (a), a detonation wave propagates combusting reactants. At location (b), an oblique shock is formed due to the pressure jump induced by a detonation wave. A slip line (c) is formed between combustion products of the current cycle and those of the previous one. Near the inflow around (d), high pressures immediately behind the detonation wave prevent inflow of fresh reactants as described in Section 2.2. As expansion occurs downstream of the detonation wave, fresh reactants are injected into the domain near (e). Near (f), Deflagration occurs along the interface between these injected reactants and the detonation products. As shown in Figure 13, the pressure increases significantly across the detonation wave (region (a)), while no noticeable pressure change is observed across the deflagration wave (region (f)).

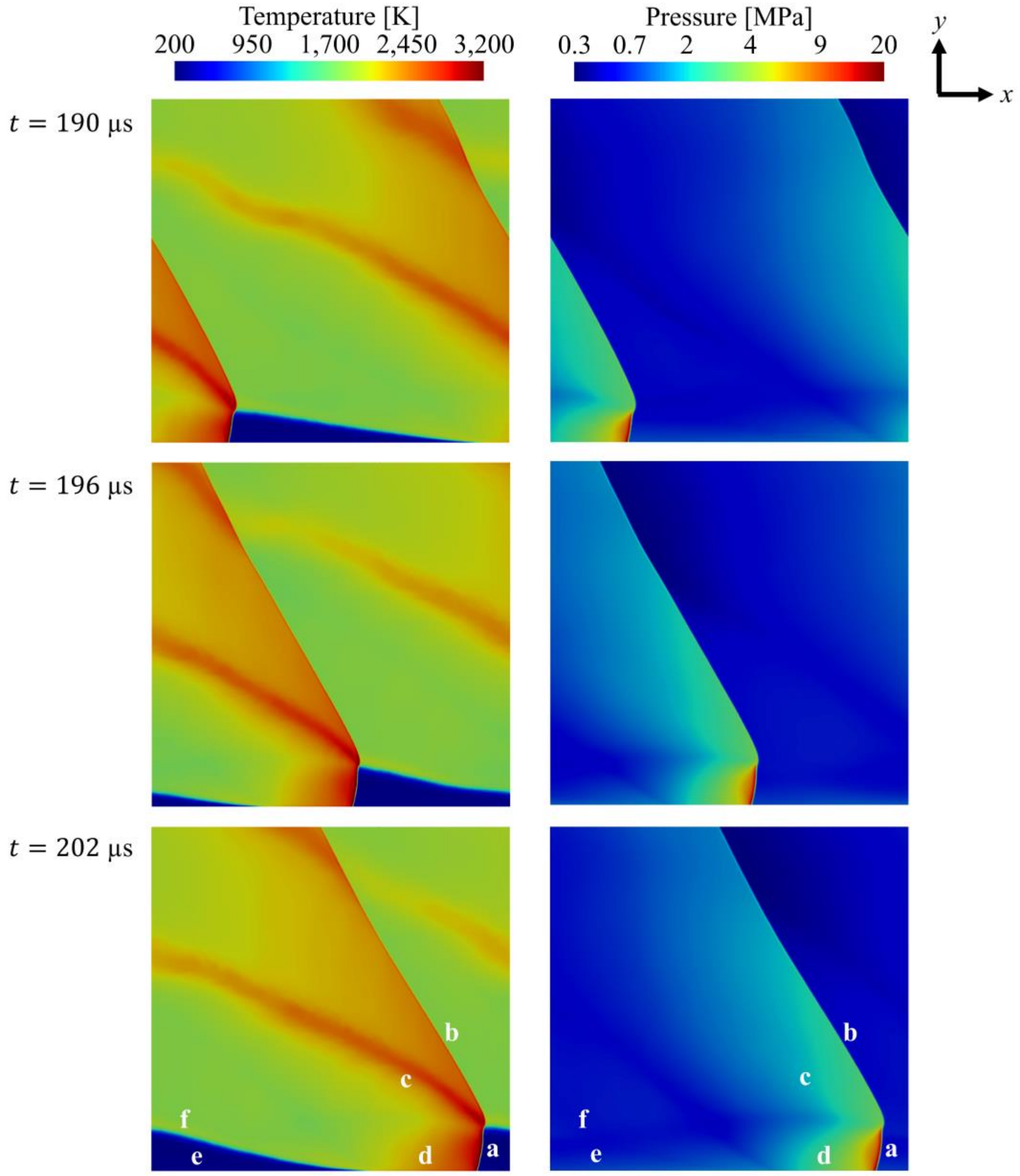

Figure 13. Snapshots of temperature (left) and pressure (right) fields from 190 μs to 202 μs. The temperature fields are presented on a linear scale, while the pressure fields are presented on a log scale.

The present simulation is validated using the results of Liu et al. [36]. Figure 14 compares temperature fields using the same color scale from Liu et al. [36]. Since Liu et al. performed simulations on a domain twice as large in the azimuthal direction, the present simulation's field is copied twice for easy comparison. The present simulation's field shows good agreement with the fields from Liu et al. [36]. The peak temperatures show similar values in both fields. Furthermore, the overall angles, which appear on detonation fronts, oblique shocks, and slip lines, and the aspect that the reactants are injected match well. The field from Liu et al. exhibits more wakes on the slip line than that from the present simulation. The upwind that leads to more numerical diffusivity induces this difference.

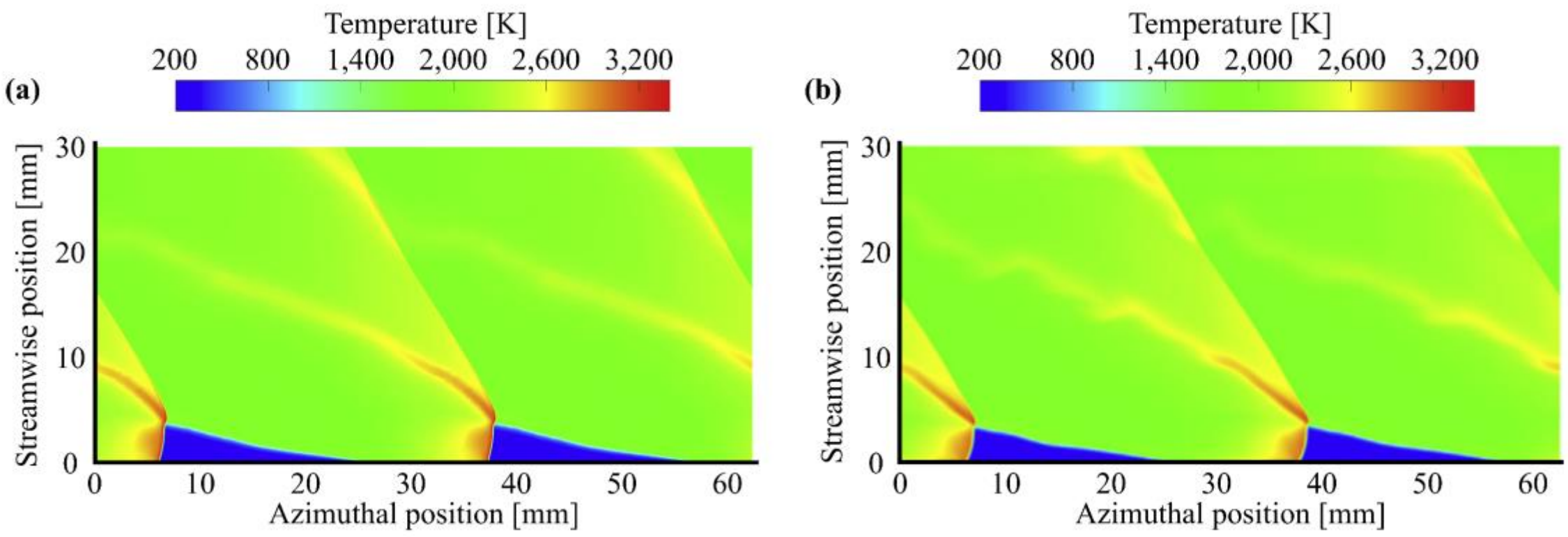

Figure 14. Comparison of temperature fields of (a) the present simulation and (b) Liu et al. [36]. Reproduced with permission from Liu et al. [36]. Copyright 2020 Elsevier.

In addition to the temperature distribution, Liu et al. reported 4 physical properties: the horizontal component of detonation velocity ($D_w$), area-averaged mass flow rate ($\dot{m}$), the specific impulse ($I_{sp}$), and the specific thrust ($F_{sp}$). The horizontal component of detonation velocity ($D_w$) is obtained by fitting linear regressions to the detonation-wave position versus time at the bottom and using each slope as the corresponding velocity. Other properties are calculated as follows [36]:

$$\dot{m} = \frac{1}{L_x} \int_{inlet} \left[ \rho v \right] dx \, , \tag{9}$$

$$I_{sp} = \frac{1}{\dot{m}_{H_2} \cdot g \cdot L_x} \int_{outlet} \left[ \rho v^2 + p - p_\infty \right] dx \, , \tag{10}$$

$$F_{sp} = \frac{1}{\dot{m}_{air} \cdot L_x} \int_{outlet} \left[ \rho v^2 + p - p_\infty \right] dx \, , \tag{11}$$

where $\rho$ is density, $v$ is the streamwise velocity, $p_\infty$ is the ambient pressure, 0.1 MPa, and $Y_i$ is mass fraction of species $i$. The specific impulse represents the axial thrust normalized by the hydrogen fuel weight flow rate, whereas the specific thrust represents the axial thrust normalized by the air mass flow rate. The same definitions, integration boundaries, and mass-flow normalization used by Liu et al. [36] were adopted to ensure a consistent numerical-to-numerical comparison.

Table 7 summarizes the validation results: the first column lists the properties compared, the second and third columns present the present simulation and literature values, and the fourth column shows the relative differences. Overall, the present simulation results show good agreement with those of Liu et al. [36].

Table 7. Comparison of various properties and corresponding errors.

| Properties | Present Simulation | Liu et al. [36] | Relative difference [%] |
|---|---|---|---|
| $D_w$ [m/s] | 1,818 | 1,862 | - 2.36 |
| $\dot{m}$ [kg/m$^2$/s] | 938 | 994 | - 5.63 |
| $F_{sp}$ [N·s/kg] | 1,646 | 1,703 | -3.35 |
| $I_{sp}$ [s] | 5,884 | 5,955 | -1.19 |

Note that the horizontal component of the detonation velocity ($D_w$=1,818 m/s in present simulation) differs from CJ detonation velocity (2,007 m/s) calculated from $p_{cr}$ and $T_{cr}$ using SDToolbox [50]. The reasons for this discrepancy will be discussed in Section 3.4.3.

Next, Figure 15 shows the axial distributions of static pressure and total pressure, averaged both azimuthally and temporally. The results of the present simulation are drawn in solid red, while those of Liu et al. are illustrated in blue squares. For reference, the inlet stagnation pressure ($p_{0,inlet}$) of 1.0 MPa is added in dotted lines. Near the inlet ($y$ < 3 mm), the detonation wave compresses combustion products, leading to larger static and total pressure than $p_{0,inlet}$. Subsequently, expansion occurs, and the static pressure decreases along axial position below $p_{0,inlet}$.

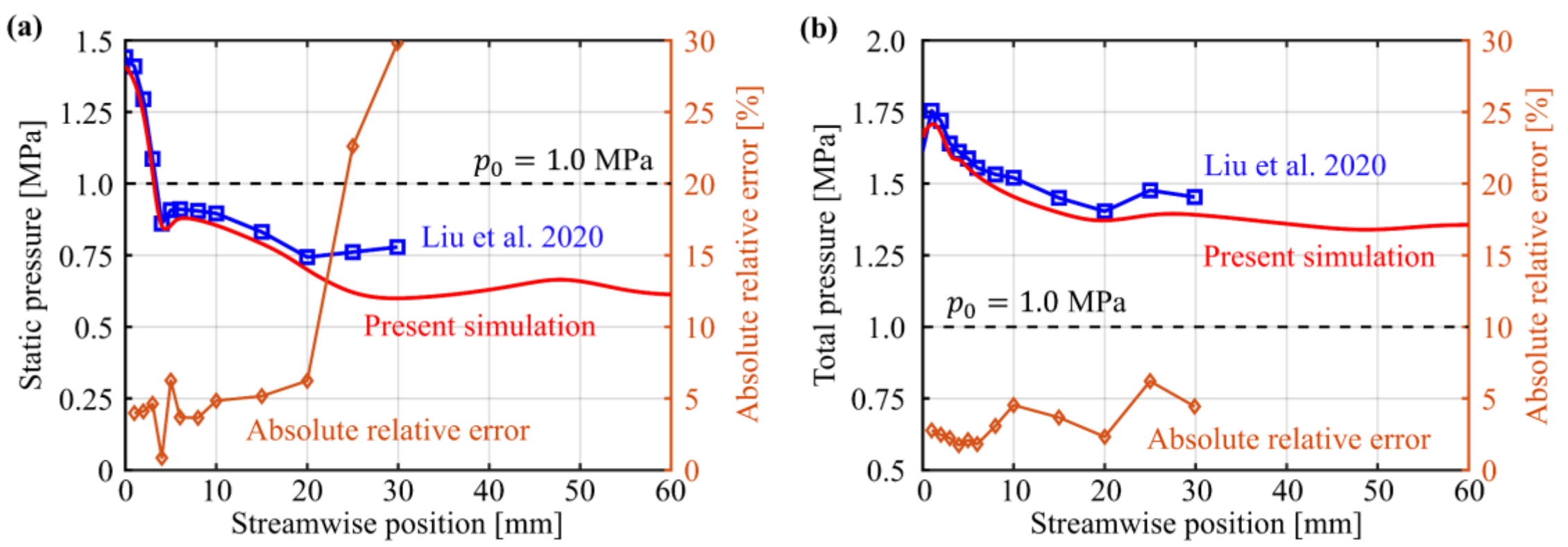

Figure 15. Axial distributions and corresponding absolute relative errors of (a) static pressure and (b) total pressure with the inlet stagnation pressure, $p_0$.

As shown in Figure 15 (a) and (b), the static and total pressures show good agreements (around 5 % relative error) up to an axial position of 20 mm. Beyond 20 mm, the difference increases significantly, which may arise from outlet boundary conditions. Note that our simulations were intentionally run in a longer domain to minimize the effects by the outlet boundary condition. Beyond the axial position of 20 mm, the total pressure stops decreasing and matches relatively well with Liu et al. [36].

The reason for different trends between static and total pressure distributions beyond 20 mm can be inferred as the factor that as expansion occurs near the outlet, the flow is accelerated, compensating for total pressure. Figure 16 shows Mach number along axial position, and Mach number increases in the region beyond the axial position of 20 mm, as we mentioned above. Also, total pressure remains higher than $p_0$, meaning that the total pressure gain at the outlet (see Figure 15 (b)). Pressure gain (PG) can be calculated by the ratio of total pressure at outlet to $p_{0,inlet}$.

$$\mathrm{PG} = \frac{p_{0,outlet} - p_{0,inlet}}{p_{0,inlet}} . \tag{12}$$

The pressure gain of the present simulation is 36.82%, which is 8.59 percentage points lower than the 45.41% reported by Liu et al. [36].

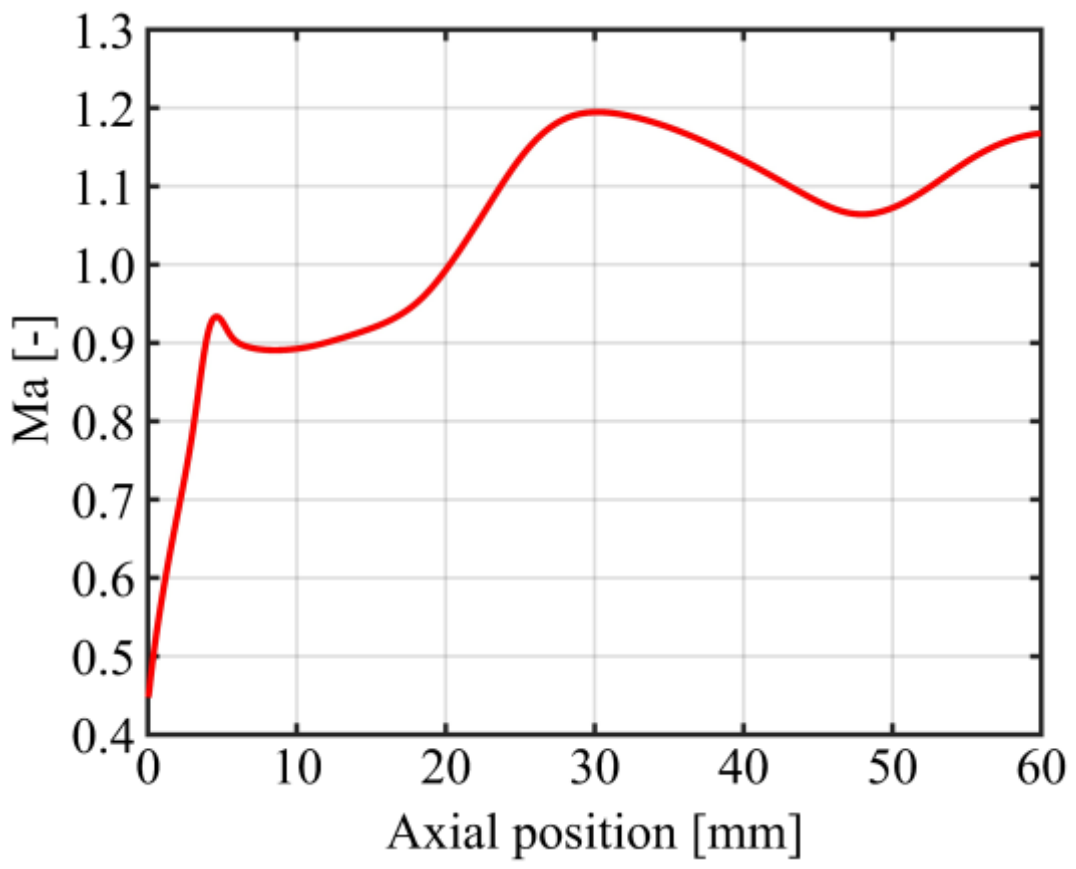

Figure 16. Axial distribution of cross-stream averaged Mach number.

*3.4.3. Analysis on horizontal components of detonation velocity in RDE*

In Section 3.4.2, the horizontal components of detonation velocity ($D_w$) from the present simulation and Liu et al. are comparable by 2.36 % as shown in Table 7. However, $D_w$ of the current simulation is lower than the CJ detonation velocity ($D_{CJ}$) by 9.42 % as mentioned above. The horizontal components of detonation velocity in RDE are lower than CJ speed due to (*i*) the detonation front angle and (*ii*) the stream tube expansion as shown in Figure 17.

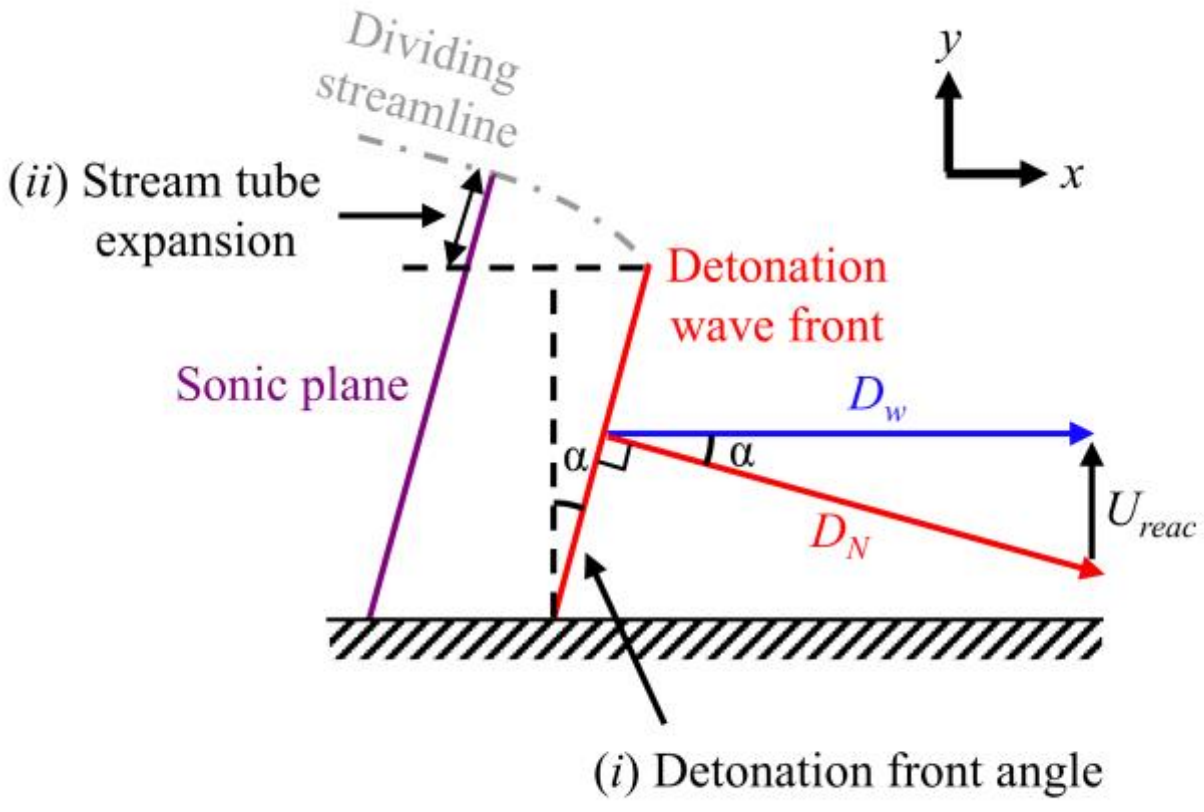

Figure 17. Brief description of a flow field in RDE with velocity vectors regarding a detonation wave.

First, we need to consider the detonation front angle to compensate for the horizontal components of detonation velocity. Figure 17 roughly displays three following velocity vectors: the horizontal components of detonation velocity ($D_w$), the reactant velocity ($U_{reac}$), and the normal detonation velocity ($D_N$). Although detonation waves in RDE seem to move horizontally along vector $D_w$, detonation fronts are tilted due to reactant inflow ($U_{reac}$). Thus, the normal detonation velocity ($D_N$) appears to be downward and can be assessed by summing two vectors, $D_w$ and $U_{reac}$. The summation of these vectors can be simplified using the detonation front's slope if the velocity of reactants only has the vertical component.

Figure 18 exhibits the horizontal velocity component ($U_x$) at 196 μs. Figure 18 (a) shows the $U_x$ field while Figure 18 (b) plots the $U_x$ values along the horizontal line at the height with the arrows. In Figure 18 (a), the $U_x$ values in the region that the reactants enter are approximately 0. Specifically, Figure 18 (b) ensures that there is no movement of reactants along azimuthal direction ($x$) in front of the detonation wave, which means that the velocity of reactants only has the vertical component. Thus, the relation between the normal detonation velocity ($D_N$) and its horizontal component ($D_w$) can be assessed using the tilted angle of detonation fronts ($\alpha$), adopting the following equation.

$$D_w = D_N \times \cos\alpha \tag{13}$$

For the calculation of $\alpha$, the detonation fronts are extracted, and the regression line is drawn to calculate its slope.

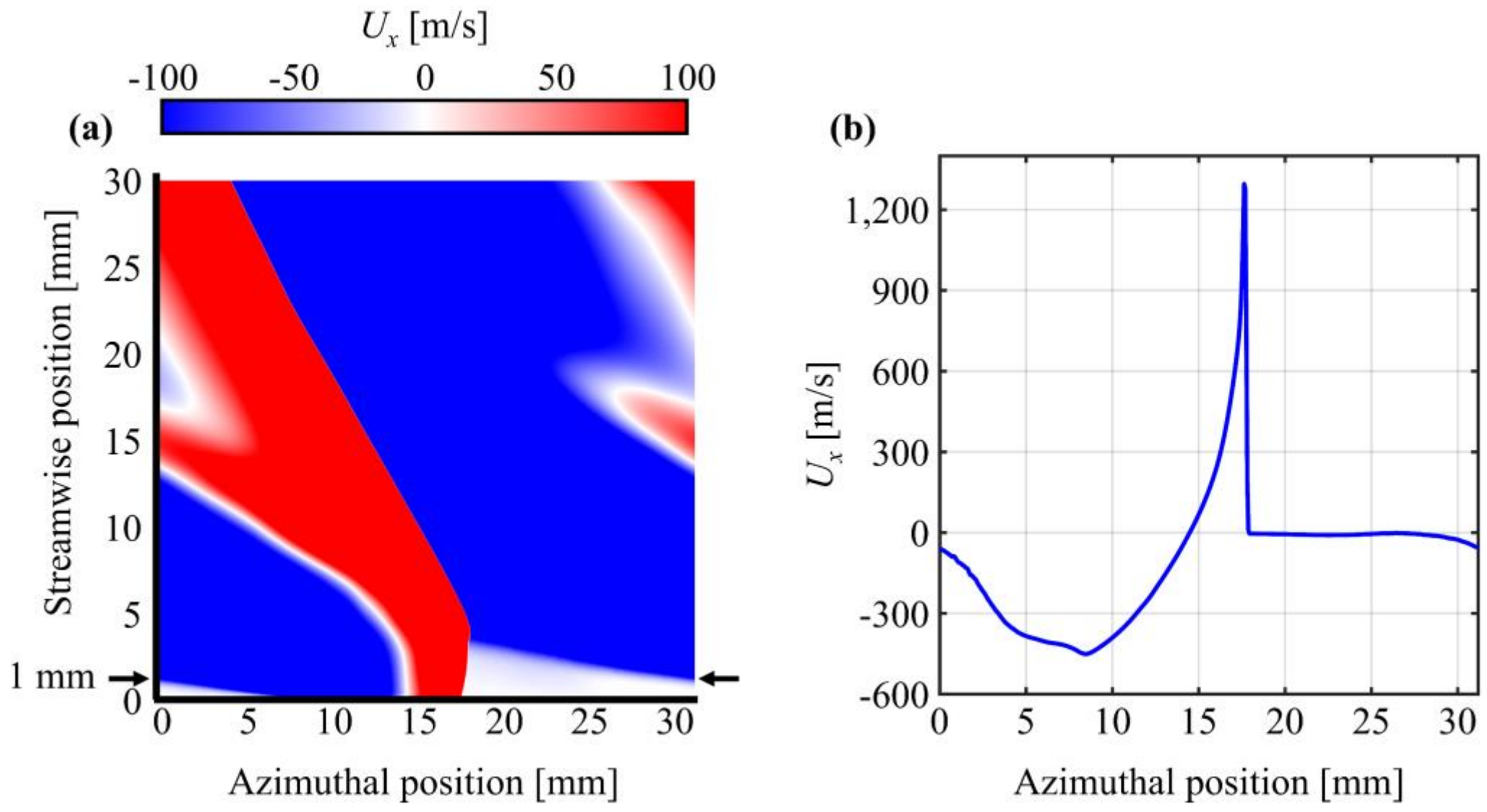

Figure 18. (a) Field of horizontal component of velocity ($U_x$), (b) $U_x$ as a function of azimuthal position at the marked height at 196 μs. The temperature and pressure field at the same time is illustrated in the second row of Figure 13.

Next, regarding the stream tube expansion, Kanda et al. [52] analyzes this velocity deficit in RDEs on ZND coordinates system in which the position of a detonation wave is fixed. The detonation velocity is reduced in RDEs since the detonation occurs under stream tube expansion conditions. They derived an equation for reduced detonation velocity as a function of $\eta_{ws}$ ($=w_2/w_1$, the ratio of the sonic point width ($w_2$) to the detonation front width ($w_1$)). Then, Mach number of the reduced detonation velocity ($M_{det}$) can be calculated by the following quartic equation as:

$$AM_{det}{}^4 - 2BM_{det}{}^2 + 1 = 0\,, \tag{14}$$

where the constants $A$ and $B$ are determined by:

$$A = \gamma_m^{\ 2} - \frac{(\gamma_c \cdot \eta_{ws} + 1)^2}{(\gamma_c + 1)\eta_{ws}^{\ 2}} \cdot \frac{\gamma_m^{\ 2}}{\gamma_c^{\ 2}} \cdot (\gamma_c - 1), \tag{15}$$

$$B = \frac{(\gamma_c \cdot \eta_{ws} + 1)^2}{(\gamma_c + 1)\eta_{ws}^{\ 2}} \cdot \left\{ \frac{\gamma_m^{\ 2}}{\gamma_c^{\ 2}} \cdot \frac{(\gamma_c - 1)}{(\gamma_m - 1)} + \frac{\gamma_m R_c}{\gamma_c R_m} \cdot \frac{q}{c_{p,c} \cdot T_1} \right\} - \gamma_m, \tag{16}$$

where $\gamma$ is the specific heat ratio, $R$ is the gas constant, and $q$ is the heat release in detonation combustion. The subscripts, $m$ and $c$, indicate the states of reactants and products, respectively. $c_{p,c}$ is the specific heat at constant pressure and $T_1$ is the temperature of reactants.

The ratio of Mach number of the reduced detonation velocity ($M_{det}$) to Mach number of CJ detonation velocity can be calculated as a function of $\eta_{ws}$, given properties calculated by SDToolbox [50]. If the sonic point width is same as the detonation front width ($\eta_{ws}$=1, a typical 1D case), $M_{det}$ is same as $M_{CJ}$. As the sonic point width is larger than the detonation front width ($\eta_{ws}$>1), $M_{det}/M_{CJ}$ is lower than the unity, reconfirming the occurrence of the velocity deficit. We calculated $M_{det}/M_{CJ}$ of the present simulation with the obtained $\eta_{ws}$ for more accurate validation of the result of the present simulation. Also, we performed coordinates translation to ZND system for subsequent analyses.

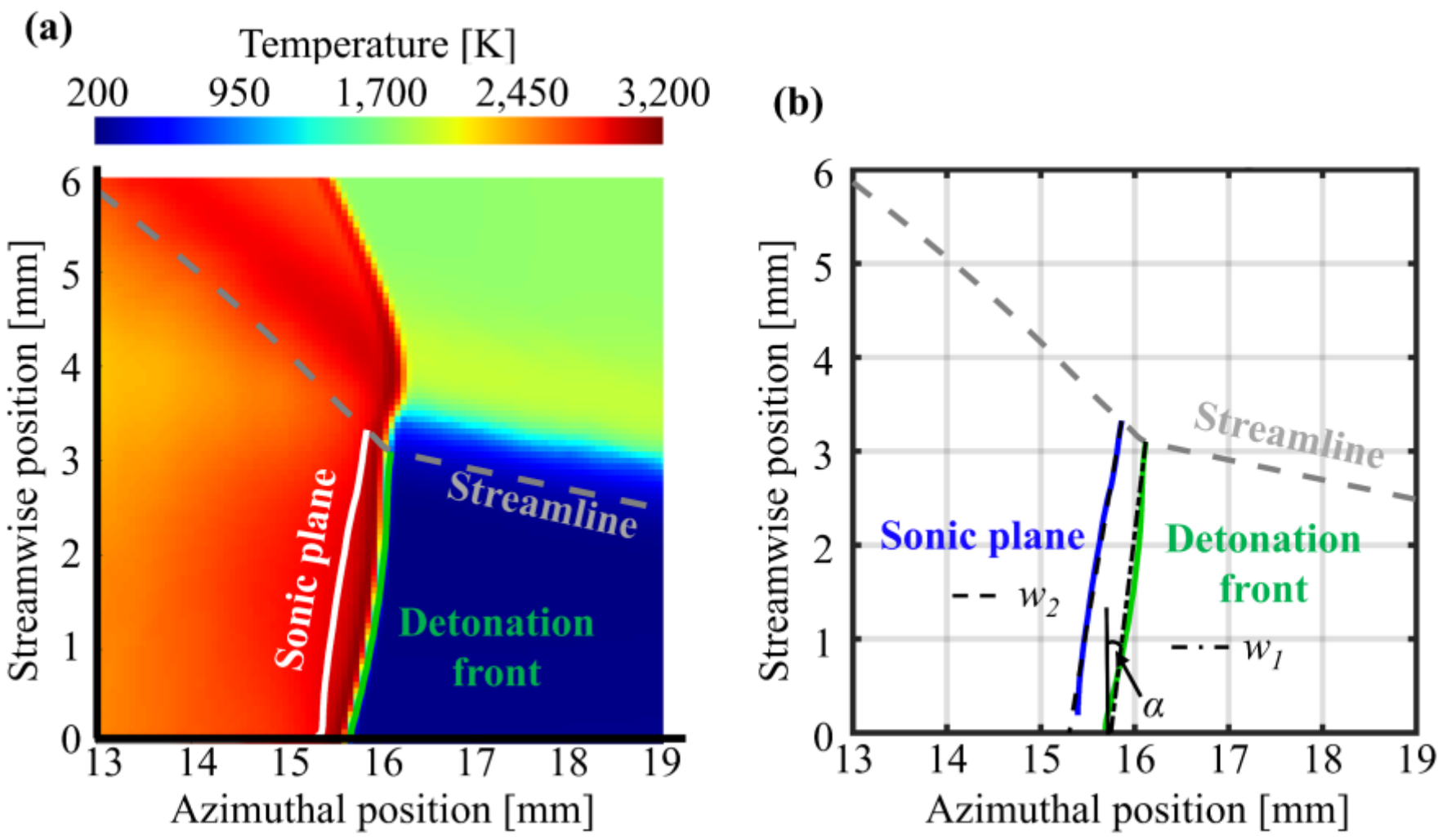

Figure 19. (a) illustrates sonic plane (white), detonation front (green), and streamline (gray) overlaid on the temperature field at 196 μs, while (b) shows the process obtaining $w_1$ and $w_2$, and the detonation front angle. Both (a) and (b) are drawn on ZND coordinate system.

As mentioned above, $w_1$ and $w_2$ are required for the evaluation of the correction factor $M_{det}/M_{CJ}$. For clarity,

Figure 19 (a) shows the sonic plane, streamline, and detonation front on the temperature field at the specified time step. In Figure 19 (b), $w_1$ is extracted from the regression line of the detonation fronts assessed by the surface where temperature is 1,600 K, and $w_2$ is determined using the regression lines fitted to the sonic plane data below the streamline originating from the highest point of the detonation front. The evaluated values of $\alpha$, $w_1$, $w_2$, $\eta_{ws}$, and $M_{det}/M_{CJ}$ are summarized in Table 8.

Table 8. Average values of various properties used for correction factor calculations.

| | $\alpha$ | $w_1$ | $w_2$ | $\eta_{ws}$ (= $w_2/w_1$) | $M_{det}/M_{CJ}$ |
|---|---|---|---|---|---|
| Values | 8.62 ° | 3.27 mm | 3.67 mm | 1.12 | 0.928 |

The correction factor for the unconfined boundary condition can be adopted to the CJ detonation velocity as the following equation.

$$D_{CJ,corr} = D_{CJ} \times \left( M_{det} / M_{CJ} \right) \times \cos\alpha \tag{17}$$

Thus, the corrected CJ detonation velocity ($D_{CJ,corr}$) and $D_w$ are presented in Table 9 with the corresponding error. The correction process reduces the relative error in terms of the detonation velocity, improving the credibility of the present simulation.

Table 9. Comparison of corrected CJ detonation velocity ($D_{CJ,corr}$) and horizontal components of detonation velocity ($D_w$).

| | Speed [m/s] | Relative error [%] |
|---|---|---|
| $D_{CJ,corr}$ | 1,841 | - |
| $D_w$ | 1,818 | - 1.27 |

The sensitivity of the corrected CJ velocity to the temperature isotherm used for detonation-front detection was examined using thresholds of 1500, 1600, and 1700 K. The same simulation dataset was used for all three thresholds; therefore, the comparison represents the sensitivity to the front-detection criterion rather than differences between simulations. As the isotherm threshold increased from 1500 to 1700 K, $M_{det}/M_{CJ}$ decreased from 0.935 to 0.920, and $D_{CJ,corr}$ decreased from 1,855 to 1,826 m/s. The corresponding relative difference between $D_w$ and $D_{CJ,corr}$ varied from −2.01% to −0.42%. Thus, although the calculated correction exhibits some dependence on the selected temperature isotherm, the corrected CJ velocities remain within a relatively narrow range.

Table 10. Sensitivity of the corrected CJ velocity to the temperature isotherm threshold used for detonation-front detection. The relative error is calculated as $\frac{(D_w - D_{CJ,corr})}{D_{CJ,corr}} \times 100\%$ with $D_w$=1,818 m/s.

| Isotherm threshold | $M_{det}/M_{CJ}$ | $D_{CJ,corr}$ [m/s] | Relative error [%] |
|---|---|---|---|
| 1500K | 0.935 | 1,855 | - 2.01 |
| 1600K | 0.928 | 1,841 | - 1.27 |
| 1700K | 0.920 | 1,826 | - 0.42 |

### *3.5. Optimization using AMR and DLB*

Building on the simulations from Section 3.4, this section demonstrates the successful application of optimization using AMR and DLB. We focus primarily on computational speed-up as the optimization metric, while ensuring that simulation accuracy remains high. Three cases are considered to evaluate AMR and DLB performance (see Table 11): Case Uniform uses the Section 3.4 results as a baseline, Case AMR isolates the effect of AMR, and Case AMR+DLB assesses the impact of DLB. All cases employ the same finest grid resolution to ensure a fair comparison of accuracy.

#### *3.5.1. Numerical setup*

To directly apply mesh adaptation to the two-dimensional RDE configuration, the two-dimensional AMR framework implemented in the present study was employed for the AMR and AMR+DLB cases. All three cases used the identical computational domain, initial conditions, and boundary conditions described in Section 3.4.1. The Uniform case employed a uniform grid with a cell size of 0.0625 mm, whereas the AMR and AMR+DLB cases used a base grid size of 0.25 mm with up to two additional refinement levels, resulting in the same finest grid size of 0.0625 mm. This ensures a consistent comparison of accuracy and computational performance among the three cases. The AMR settings are summarized in Table 12. Cells satisfying the refinement criterion were refined to capture the key RDE flow structures, including the detonation front, oblique shocks, and deflagration region.

Table 11. Description of test cases.

| Case name | AMR | DLB | Smallest grid size |
|---|---|---|---|
| Uniform | X | X | |
| AMR | O | X | 0.0625 mm |
| AMR+DLB | O | O | |

Table 12. Summary of AMR conditions.

| Conditions | Values |
|---|---|
| Domain size | $31.25 \text{ mm} \times 60 \text{ mm}$ |
| Maximum refinement level | 2 |
| Base grid size | 0.25 mm |
| Finest grid size | 0.0625 mm |
| Refine criterion for Level 2 | $\|\nabla\rho\| > 500$ kg/m$^4$ |

DLB redistributes the mesh dynamically during runtime according to processor load (*cpuLoad*) measurements. *cpuLoad* is assessed from the computation time each processor spends on its assigned subdomain. By considering computation time rather than cell count, regions with intensive chemical reaction workloads are properly accounted for, enabling more efficient redistribution and reduced overall computational cost. The Zoltan partitioning library is used to execute mesh redistribution based on the collected *cpuLoad* data [53]. Mesh redistribution is triggered when the *cpuLoad* imbalance exceeds 0.1.

All benchmark simulations were performed using 48 CPU cores on two Intel Xeon Gold 6248R processors operating at 3.0 GHz. The same computational resources, numerical settings, and parallel configuration were used for the Uniform, AMR, and AMR+DLB cases. The time step was adjusted dynamically to maintain a maximum Courant number of $Co_{max}$ = 0.1. This consistent configuration enables a direct comparison of the computational performance of the three cases.

*3.5.2. Simulation results*

Simulations were conducted for 500 μs and Figure 20 presents snapshots of the flow fields from 190 to 202

μs. The left column shows temperature contours, confirming accurate capture of RDE flow structures. The right column indicates mesh refinement levels: gray represents unrefined cells (Level 0), blue-gray represents cells refined once (Level 1), and dark blue represents cells refined twice (Level 2). The refinement strategy is effectively focused on the key regions of interest, including the detonation and deflagration zones, and oblique shocks. As the detonation wave propagates to the right, the refined regions move with the detonation wave.

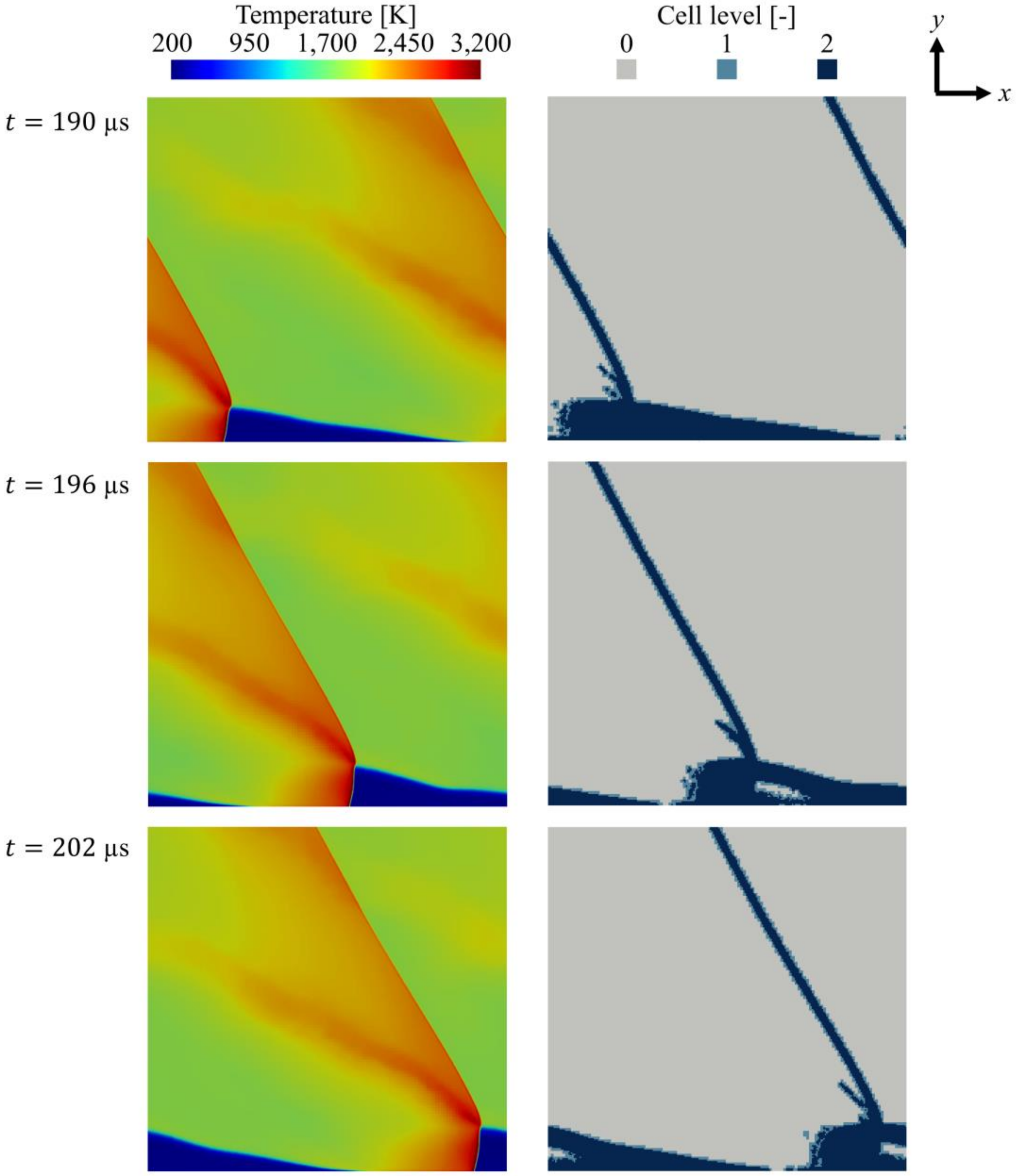

Figure 20. Snapshots of temperature (left) and cell level (right) fields of Case AMR+DLB at various time steps.

The performance metrics evaluated in Section 3.4 were calculated using the same definitions and time-averaging interval for all three cases. As summarized in Table 13, the detonation velocity, specific thrust, and specific impulse differ from the Uniform case by less than 0.5%. The area-averaged mass flux exhibits a slightly larger relative difference of 1.07%, which may be attributed to differences in the near-inlet grid structure. The AMR and AMR+DLB results are also in close agreement, although small differences are observed in some time-averaged quantities. Dynamic load balancing redistributes computational workloads among processors and modifies the associated parallel communication pattern [54, 55]. Although this redistribution does not alter the governing equations, mesh resolution, or numerical schemes, it may change the ordering of finite-precision parallel operations. Because floating-point arithmetic is non-associative, changes in the operation sequence can introduce small numerical differences [56]. Such numerical differences may accumulate over a long transient simulation and lead to minor variations in the time-averaged quantities. Therefore, the small differences between the AMR and AMR+DLB cases are attributed to numerical variations associated with dynamic mesh redistribution rather than changes in the underlying physical or numerical models. Nevertheless, the overall agreement confirms that the application of AMR and DLB has only a minor influence on the predicted RDE performance.

Table 13. Comparison of time-averaged performance metrics for the Uniform, AMR, AMR+DLB cases over t = 100-500 μs. Relative differences are calculated with respect to the Uniform case.

| Case | Properties | Values | Relative error [%] |
|---|---|---|---|
| Uniform | $D_w$ [m/s] | 1,818 | - |
| | $\dot{m}$ [kg/(m$^2$·s)] | 938 | - |
| | $F_{sp}$ [N·s/kg] | 1,646 | - |
| | $I_{sp}$ [s] | 5,884 | - |
| AMR | $D_w$ [m/s] | 1,818 | 0.00 |
| | $\dot{m}$ [kg/(m$^2$·s)] | 948 | 1.07 |
| | $F_{sp}$ [N·s/kg] | 1,638 | - 0.49 |
| | $I_{sp}$ [s] | 5,858 | - 0.44 |
| AMR+DLB | $D_w$ [m/s] | 1,818 | 0.00 |
| | $\dot{m}$ [kg/(m$^2$·s)] | 948 | 1.07 |
| | $F_{sp}$ [N·s/kg] | 1,639 | - 0.43 |
| | $I_{sp}$ [s] | 5,859 | - 0.42 |

Figure 21 compares the axial pressure distributions for Case Uniform, AMR, and AMR+DLB. Overall, Case

AMR and Case AMR+DLB show excellent agreement with Case Uniform. In particular, the pressure distributions of Cases AMR and AMR+DLB are visually indistinguishable. Toward the outlet region, small discrepancies in pressure distributions (see Figure 21 (a) and (b)) arise due to the increased proportion of coarser cells.

To further assess the accuracy of the developed two-dimensional AMR method, the time-averaged flow fields obtained from the Uniform, AMR, and AMR+DLB cases were compared in a detonation-centred coordinate system. Instantaneous solutions over t = 100–500 μs were shifted in the x-direction to align the detonation-wave position and were subsequently averaged on a common grid corresponding to the finest mesh resolution. Figure 22 compares the resulting mean temperature fields, with horizontal lines indicating the sampling locations near y = 2, 10, and 20 mm used for the subsequent profile comparisons.

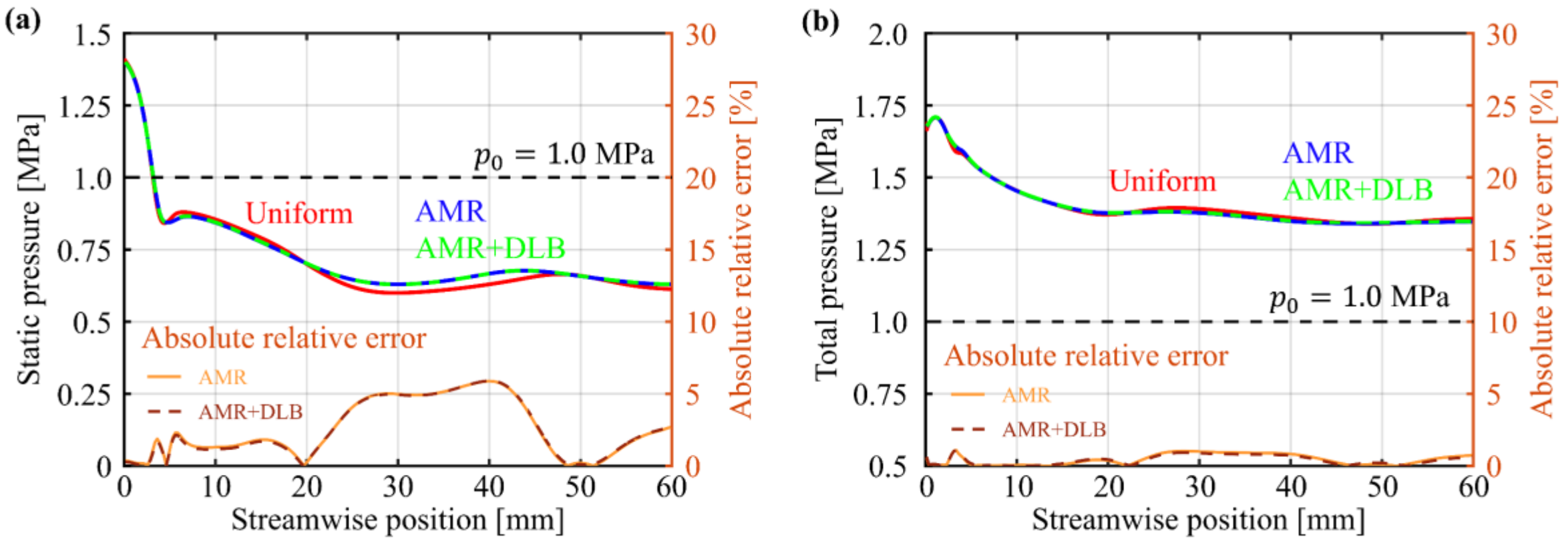

Figure 21. Axial distributions and absolute relative errors of (a) static pressure and (b) total pressure of all cases with the inlet stagnation pressure, $p_0$.

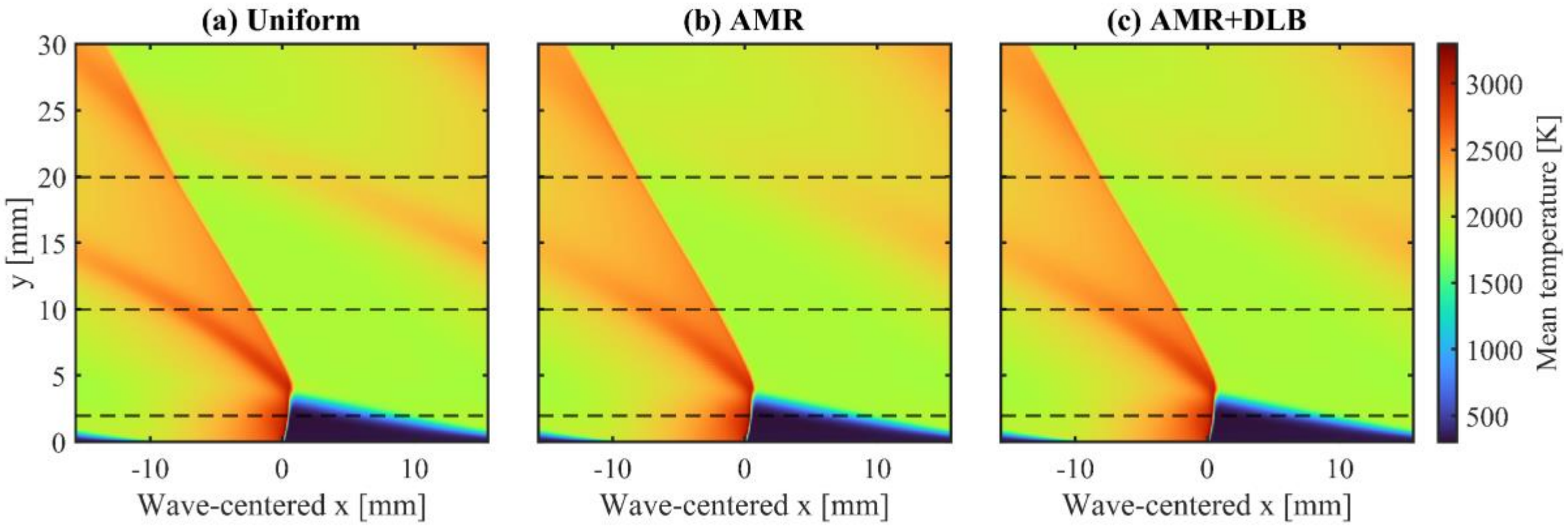

Figure 22. Detonation-centered mean temperature fields for the (a) Uniform, (b) AMR, and (c) AMR+DLB cases, averaged over t = 100–500 μs. The horizontal dashed lines indicate the nominal profile-sampling heights of y = 2, 10, and 20 mm. The same temperature range and color scale are used for all three panels.

Figure 23 and Figure 24 compare the detonation-centered mean total and static pressure profiles, respectively, at the cell-center locations closest to y = 2, 10, and 20 mm. At all three heights, the AMR and AMR+DLB profiles closely follow the Uniform results across the leading pressure rise and the subsequent expansion region. Both the locations and magnitudes of the pressure peaks are consistently reproduced, and the AMR and AMR+DLB curves remain virtually identical. These results show that the developed two-dimensional AMR method preserves the principal detonation and shock dynamics obtained using the corresponding finest uniform mesh.

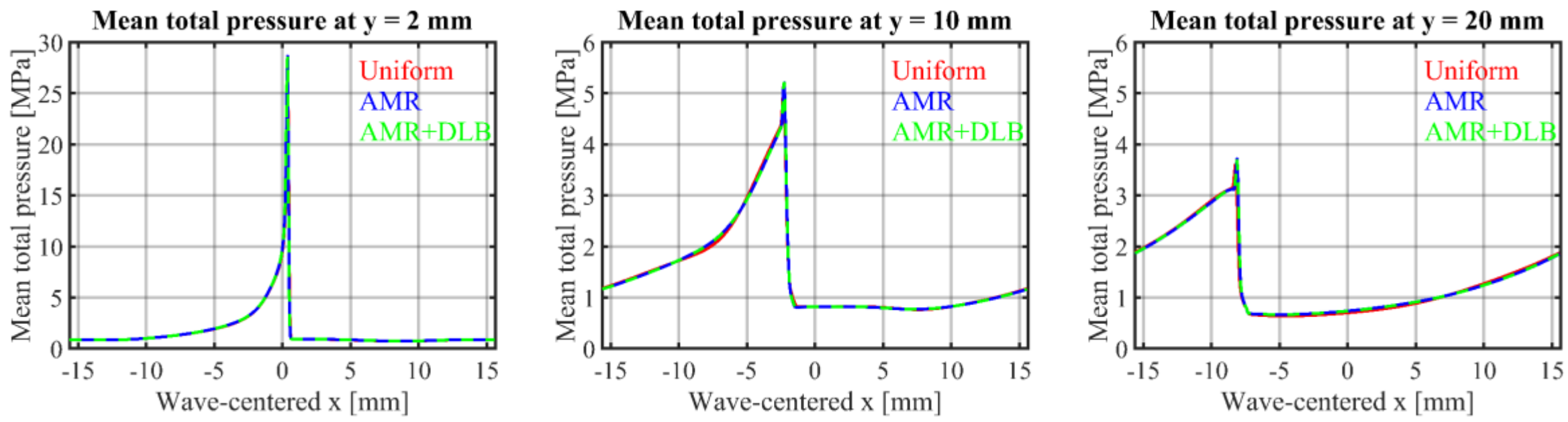

Figure 23. Detonation-centered mean total pressure profiles at the cell-center locations nearest to the nominal heights of (a) y = 2 mm, (b) y = 10 mm, and (c) y = 20 mm, averaged over $t$ = 100-500 μs. The Uniform, AMR, and AMR+DLB cases are represented by red solid, blue solid, and green dashed lines, respectively.

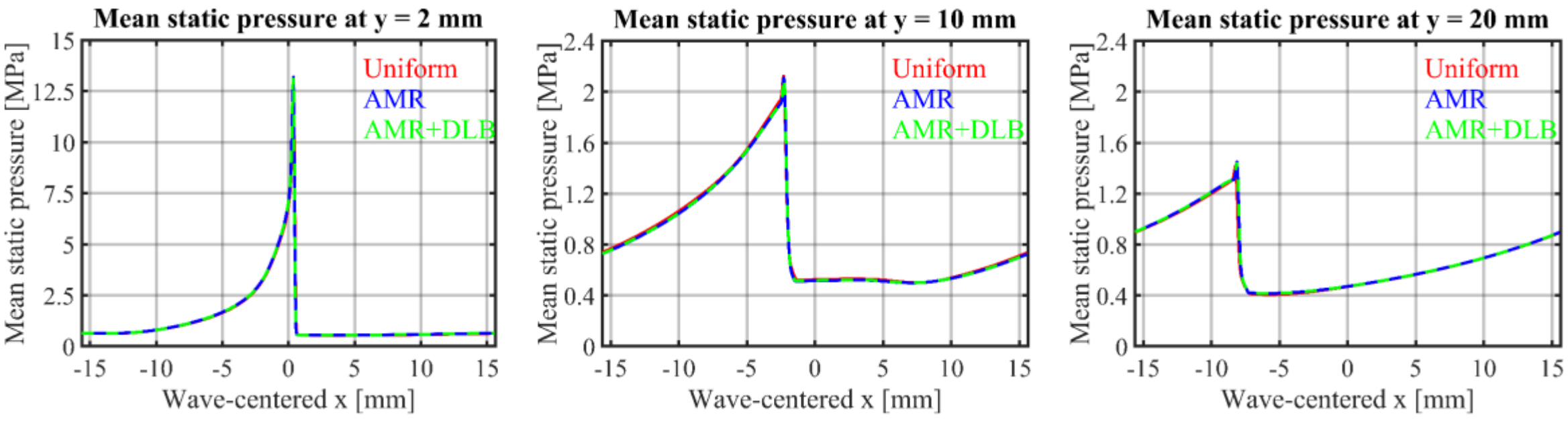

Figure 24. Detonation-centered mean static pressure profiles at the cell-center locations nearest to the nominal heights of (a) y = 2 mm, (b) y = 10 mm, and (c) y = 20 mm, averaged over $t$ = 100-500 μs. The Uniform, AMR, and AMR+DLB cases are represented by red solid, blue solid, and green dashed lines, respectively.

Figure 25 presents the corresponding mean-temperature profiles. Near y = 2 mm, where the detonation front and the adjacent reaction region are refined, the three cases exhibit close agreement in the temperature peak, the sharp post-wave decrease, and the downstream recovery. More noticeable differences appear at y = 10 and 20 mm in the slip-line region visible in Figure 22. In the Uniform case, the temperature profiles exhibit a distinct gradual rise and fall across this region, whereas the same variation is attenuated and smeared in the AMR and AMR+DLB results. The present AMR criterion is based on the density-gradient magnitude and therefore preferentially refines the detonation front, shocks, and other regions with strong density gradients. Because the thermal variation across the slip line is not accompanied by a sufficiently strong density gradient, this region largely remains at the base-grid resolution in the AMR cases. The resulting increase in numerical diffusion across the coarser cells smooths the local temperature variation and reduces its amplitude. The nearly identical AMR and AMR+DLB profiles further indicate that this difference originates from the spatial refinement criterion rather than from dynamic load balancing.

Overall, the close agreement of the total and static pressure distributions demonstrates that the developed two-dimensional AMR method accurately preserves the principal detonation and shock dynamics obtained with the finest Uniform mesh. The temperature comparison also clarifies a limitation of the density-gradient-based refinement criterion: thermal variations along the slip line can be under-resolved when the associated density gradient is weak. Furthermore, the nearly identical AMR and AMR+DLB results confirm that coupling the developed AMR method with the existing DLB functionality does not introduce a noticeable alteration of the computed flow field.

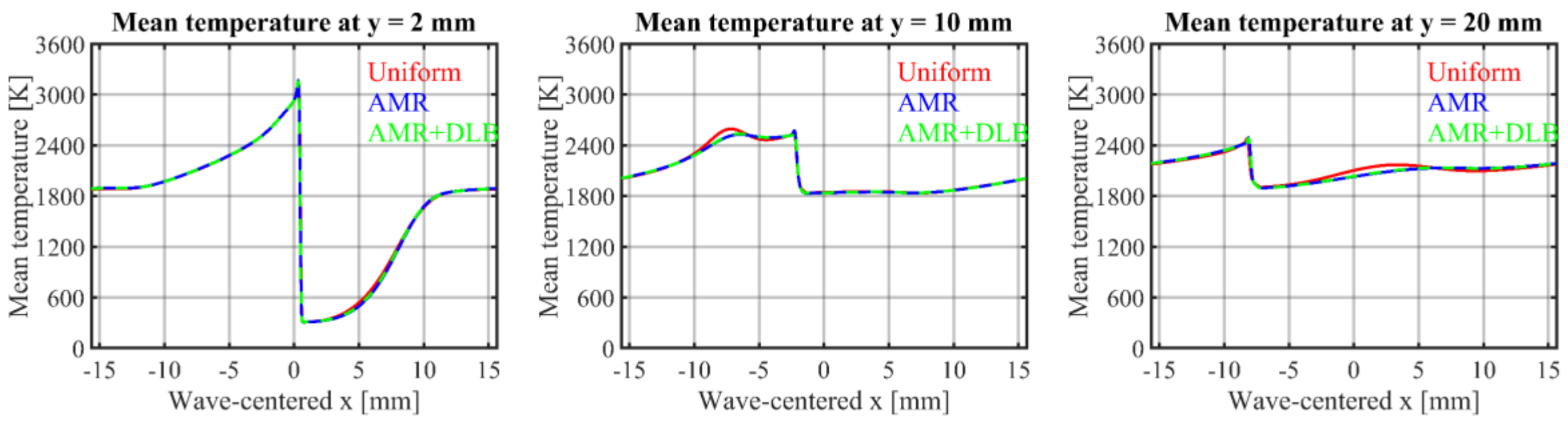

Figure 25. Detonation-centered mean temperature profiles at the cell-center locations nearest to the nominal heights of (a) y = 2 mm, (b) y = 10 mm, and (c) y = 20 mm, averaged over *t* = 100-500 μs. The Uniform, AMR, and AMR+DLB cases are represented by red solid, blue solid, and green dashed lines, respectively.

Table 14 summarizes the area occupied by each refinement level in the AMR+DLB case at *t* = 200 μs. The number of cells at each level was multiplied by the corresponding cell area to determine its occupied area and fraction of the total computational domain. Level 0 cells occupy 92.12% of the domain, whereas Level 1 and Level 2 cells account for 1.73% and 6.15%, respectively. Thus, the finest cells are concentrated within a small portion of the domain containing the detonation front, deflagration region, and oblique shocks, demonstrating the effectiveness of AMR in reducing the computational cost of the RDE simulation.

Table 14. Area fractions of each refinement level at t = 200 μs for Case AMR+DLB.

| Cell Level | No. of cells | Area of each cell | Occupied area | Area fraction |
|---|---|---|---|---|
| 0 | 27,635 | $(0.25\ mm)^2$ | 1,727.188 $mm^2$ | 92.12 % |
| 1 | 2,081 | $(0.125\ mm)^2$ | 32.516 $mm^2$ | 1.73 % |
| 2 | 29,516 | $(0.0625\ mm)^2$ | 115.297 $mm^2$ | 6.15 % |
| Total | 59,232 | - | 1,875.000 $mm^2$ | 100.0 % |

Table 15 compares the computational costs of the three cases. The wall-clock time was measured while advancing each simulation from 200 to 201 μs using 48 CPU cores. The reported wall-clock time includes the computational costs of flow-field solution, chemical-kinetics integration, mesh refinement, and dynamic mesh redistribution incurred during the measured interval. Field-data output was included in the reported wall-clock time. Applying AMR (Case AMR) reduced the overall cell count dramatically and accelerated computation by 1.5 times. When DLB was added to AMR (Case AMR+DLB), computation time fell even further, achieving a 4.2-fold speed-up over Case Uniform and a 2.8-fold speed-up over Case AMR. These results demonstrate that DLB is essential to fully exploit AMR's cost-optimization benefits.

Further analyses were performed to assess the effect of DLB. Figure 26 (a) and (b) show the processor-subdomain configurations for the Uniform and AMR cases, respectively, at $t$ = 200 μs. Both cases use a fixed decomposition consisting of six subdomains in the x-direction and eight in the y-direction. In the AMR case, this fixed decomposition does not account for the spatially nonuniform distribution of cells and computational workload, resulting in significant load imbalance. Figure 26 (c) and (d) show the dynamically redistributed subdomains for the AMR+DLB case at $t$ = 200 and 201 μs, respectively. As the detonation wave propagates to the right, the region containing densely distributed subdomains moves with the wave, thereby maintaining a more balanced computational workload among the processors.

Table 15. Comparison of cell counts at 200 μs, wall-clock times required to advance the solution from 200 to 201 μs, and speedups relative to Case Uniform.

| Case | No. of cells | Wall-clock time (200 to 201 μs) | Speed-up (relative to Uniform) |
|---|---|---|---|
| Uniform | 480,000 | 281 s | - |
| AMR | 59,481 | 188 s | 1.5 |
| AMR+DLB | 59,232 | 67 s | 4.2 |

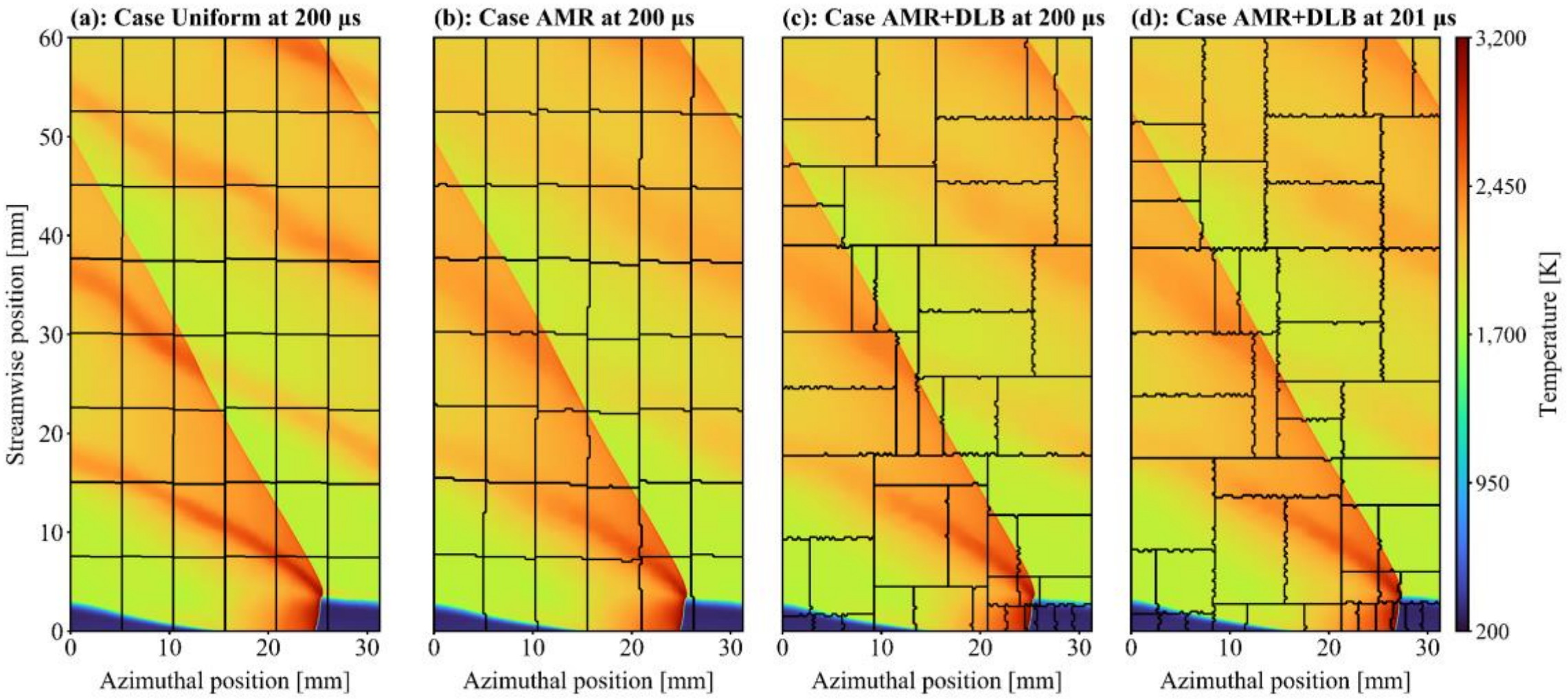

Figure 26. Subdomain compositions over temperature fields for (a) Case Uniform at 200 μs, (b) Case AMR at 200 μs, (c) Case AMR+DLB at 200 μs, and (d) Case AMR+DLB at 201 μs.

Figure 27 shows exemplary distributions of the number of cells on each thread at $t$ = 200 μs. Although the figure presents the results at a specific time, similar trends are observed throughout the simulation. The horizontal axis represents thread IDs (from 1 to 48), and the vertical axis indicates the number of cells assigned to each thread. In Figure 27 (a) (Case Uniform), each thread is assigned approximately 10,000 cells, indicating a balanced workload across all threads. In Figure 27 (b) (Case AMR), however, some threads contain a large number of cells (up to 6,816 cells), while most threads are assigned approximately 600 to 700 cells. This leads to an overload on specific threads, causing bottleneck effects and increasing the overall computational cost. Furthermore, as the detonation propagates in Case AMR, changes are observed in the thread IDs where cell concentration occurs. Figure 27 (c) (Case AMR+DLB) shows that DLB redistributes the AMR mesh much more evenly, with 800 to 1,324 cells assigned per thread. Compared to Case AMR, cells concentrated on a small number of threads (as shown in Figure 27 (b)) are more evenly distributed in Case AMR+DLB, which helps eliminate bottlenecks and accelerate computation.

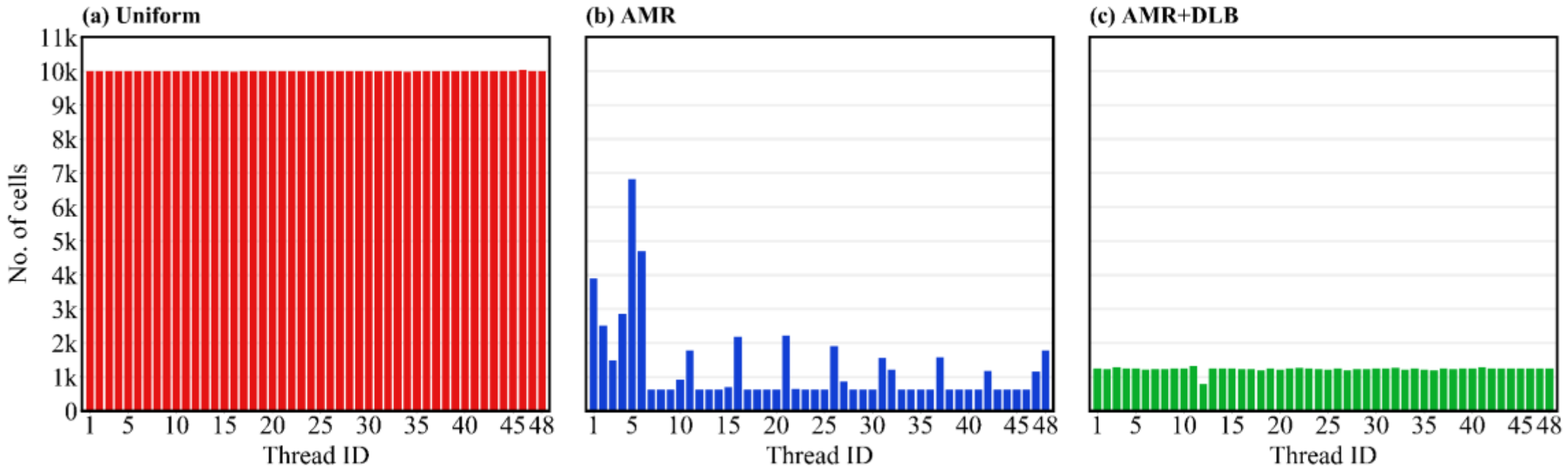

Figure 27. Distributions of the number of cells assigned to each thread for (a) Uniform, (b) AMR, and (c) AMR+DLB cases at $t$ = 200 µs.

## 4. Conclusion

In this work, an open-source framework for RDE simulations was developed using OpenFOAM v12 and accelerated through AMR and DLB. The *multicomponentFluid* solver was first validated using a one-dimensional detonation-tube problem. The calculated detonation velocities agreed with the CJ velocity to within 0.1%, and the result obtained using a grid spacing of 0.1 mm agreed closely with the experimental data of Malik et al. [48]. The corresponding differences from the CJ velocity and the experimental result were -0.10% and 0.06%, respectively.

A separate grid-convergence study was conducted to assess the capability of the pressure-based solver to reproduce the detailed ZND structure. As the grid was refined, the temperature, pressure, density, and species profiles progressively approached the SDToolbox ZND solution. The induction-zone length was quantitatively resolved at grid spacings of 0.5 µm or smaller, for which the difference from the ZND value remained below 1%. At the finest grid spacing of 0.015625 µm, the predicted induction length differed from the ZND value by 0.18%. Nevertheless, a pressure overshoot remained immediately behind the leading shock at all grid resolutions, indicating a limitation of the pressure-based formulation in capturing strong discontinuities. These results demonstrate that the induction-zone structure can be resolved using sufficiently fine grids. However, the demanding resolution requirements impose a substantial computational burden, limiting the practicality of resolving such detailed structures in RDE simulations.

The two-dimensional cellular detonation simulation was performed using a base grid spacing of 200 µm and three levels of AMR, resulting in a finest grid spacing of 25 μm. The numerical soot foil revealed a cellular propagation pattern with a representative cell size of approximately 62 mm by 30 mm, compared with 54 mm by 31 mm reported by Oran et al. [49]. The cellular pattern did not reproduce the reference result exactly, and vertical numerical traces associated with the leading-shock pressure overshoot were present in the numerical soot foil. The comparison indicates agreement in the approximate cell scale and qualitative cellular pattern, but does not establish quantitative accuracy for detailed cellular dynamics.

The two-dimensional RDE simulation framework was also validated against the numerical results of Liu et al. [36]. The temperature distribution and the principal performance quantities, including detonation velocity, area-averaged mass flux, specific thrust, specific impulse, and axial pressure distribution, showed good agreement with the reference results. The horizontal component of the calculated detonation velocity differed from the corrected CJ velocity by 1.27%.

Finally, AMR and DLB reduced the computational cost while maintaining comparable RDE performance predictions. The principal validation quantities differed by no more than 1.07% among the uniform-grid, AMR, and AMR+DLB simulations. AMR alone provided a 1.5-fold speed-up, whereas the combined use of AMR and DLB accelerated the computation by up to 4.2-fold relative to the uniform-grid simulation. These results demonstrate the computational benefits of combining AMR and DLB in the present two-dimensional RDE simulations. However, the slip-line region remained susceptible to numerical diffusion and insufficient refinement under the density-gradient-based AMR criterion. For studies targeting detailed shear-layer dynamics, higher-order discretization should be assessed to reduce numerical diffusion, together with alternative refinement criteria that better identify shear regions. Additionally, with further validation and computational performance assessment, the framework could be applied to large-scale three-dimensional RDE simulations.

Overall, the present results support the use of the framework for predicting global RDE flow characteristics and performance quantities at computationally practical grid resolutions under the conditions examined. Detailed detonation structures and fine-scale instabilities remain outside the scope of the present RDE assessment.

# Appendix

*Appendix A. Governing Equations*

The conservation equations of mass, momentum, energy, and species are given as follows:

$$\frac{\partial \rho}{\partial t} + \nabla \cdot (\rho \mathbf{u}) = 0 , \tag{18}$$

$$\frac{\partial (\rho \mathbf{u})}{\partial t} + \nabla \cdot (\rho \mathbf{u}\mathbf{u}) = -\nabla p + \nabla \cdot \boldsymbol{\tau} , \tag{19}$$

$$\frac{\partial (\rho h)}{\partial t} + \nabla \cdot (\rho \mathbf{u} h) + \frac{\partial (\rho K)}{\partial t} + \nabla \cdot (\rho \mathbf{u} K) - \frac{\partial p}{\partial t} = -\nabla \cdot \mathbf{q} + \dot{Q} , \tag{20}$$

$$\frac{\partial (\rho Y_i)}{\partial t} + \nabla \cdot (\rho \mathbf{u} Y_i) + \nabla \cdot (\rho \mathbf{V}_i Y_i) = \dot{\omega}_i , \tag{21}$$

where $t$ is time, $\rho$ is the density, $\mathbf{u}$ is the velocity vector, and $p$ is the pressure. $\boldsymbol{\tau}$ is the viscous stress defined as

$$\boldsymbol{\tau} = -\frac{2}{3}\mu(\nabla \cdot \mathbf{u})\mathbf{I} + \mu\left[\nabla \mathbf{u} + (\nabla \mathbf{u})^T\right] , \tag{22}$$

where $\mu$ is the dynamic viscosity, and $\mathbf{I}$ is the unit tensor. $h$ denotes the specific enthalpy, and $\dot{Q}$ represents the heat release from reactions. $K$ and $\mathbf{q}$ are the specific kinetic energy and the heat flux vector, and are defined as follows:

$$K = \frac{\left|\mathbf{u}^2\right|}{2} , \tag{23}$$

$$\mathbf{q} = -\frac{\lambda}{c_p}\nabla h , \tag{24}$$

where $\lambda$ is the thermal conductivity and $c_p$ is the specific heat at constant pressure. $Y_i$ is the mass fraction of species $i$, and $\dot{\omega}_i$ is the production rate of species $i$. $\boldsymbol{V}_i$ is the diffusion velocity of species $i$ given by Fick's Law:

$$\mathbf{V}_i = -\frac{D_{i,mix}}{X_i}\nabla X_i \tag{25}$$

where $D_{i,mix}$ is the mixture-averaged diffusion coefficient, and $X_i$ is the mole fraction of species $i$. In *multicomponentFluid* solver, unity Lewis number ($Le$) assumption is applied to calculate $D_{i,mix}$:

$$Le = \frac{\alpha}{D_{i,mix}} \tag{26}$$

where $\alpha$ is the thermal diffusion coefficient.

*Appendix B. Variation of the equivalence ratio at the inlet in z-direction*

To provide a preliminary demonstration of the qualitative response to three-dimensional effects, simulations were performed for an unwrapped RDE configuration with prescribed variations along the spanwise direction. Specifically, eight computational cells are assigned across the thickness, and the local equivalence ratio is prescribed to change gradually from 0.9 to 1.1 in the z-direction. Although this approach does not constitute a fully three-dimensional annular geometry, it allows us to systematically investigate how spanwise mixture inhomogeneity influences the propagation and stability of the detonation front. The present analysis provides a preliminary qualitative illustration of the response to spanwise mixture variations, motivating future assessment of the framework in fully three-dimensional RDE configurations.

The computational geometry employed in this study is identical to that described in Section 3.4. The only modification concerns the inlet boundary condition at the bottom boundary, where premixed hydrogen–air is injected. Instead of a uniform composition, the local mixture fraction is prescribed such that the equivalence ratio varies smoothly from 0.9 to 1.1 across the *z*-direction. This variation is imposed using a cosine profile, ensuring a gradual and continuous change in species mass fractions along the thickness direction.

$$\phi(z) = 1 + 0.1 \cdot \cos(4\pi z) \tag{27}$$

By introducing this controlled inhomogeneity at the inlet, the simulation enables systematic investigation of the effect of spanwise mixture variations on detonation front dynamics.

Figure 28 illustrates the inlet equivalence ratio distribution along the *z*-direction. Due to the cosine variation, the mixture is richest ($\phi$ = 1.1) at the boundaries, $z$ = 0 and 0.5 mm, and leanest ($\phi$ = 0.9) at the midpoint ($z$ = 0.25 mm). This results in a hydrogen mass fraction that follows a shape similar to a cosine curve, being highest at the boundaries and lowest in the center. Conversely, the air mass fraction exhibits the opposite trend, forming a cosine curve that is lowest at the boundaries and highest at the center.

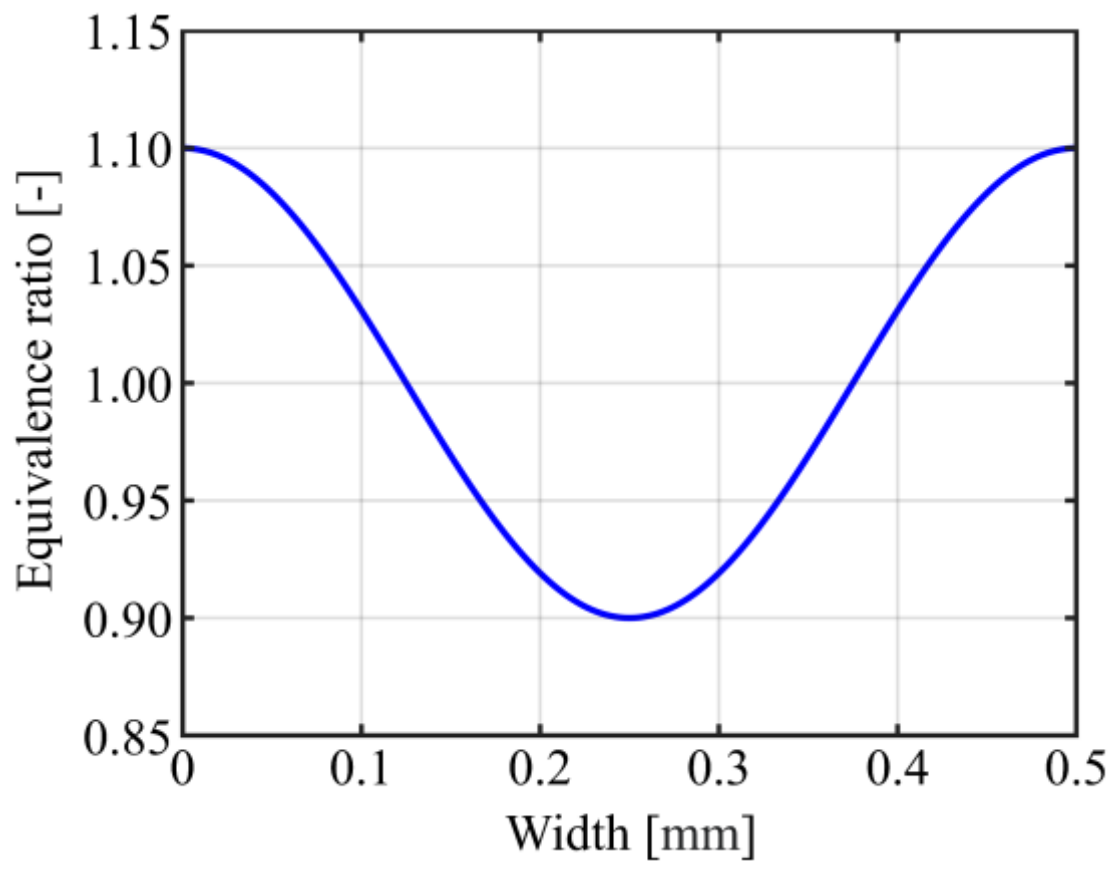

Figure 28. Spanwise variation of inlet equivalence ratio. The equivalence ratio varies from $\phi$ = 0.9 at the midplane to $\phi$ = 1.1

near the boundaries following a cosine profile.

We analyze the impact of the equivalence ratio variation along the z-direction on the detonation front. As previously described, the equivalence ratio smoothly transitions from 0.9 at the center to 1.1 at the boundaries, introducing spanwise mixture inhomogeneity. This gradient is expected to influence the detonation front, affecting both its structure and propagation velocity. A detailed examination of how these local variations in equivalence ratio influence the front's characteristics will be provided.

Figure 29 (a) presents the temperature field over the unwrapped RDE with the equivalence ratio variation along the z-direction. Despite this spanwise (z-direction) equivalence-ratio variation, the RDE maintained stable operation, with the detonation front propagating steadily. Figure 29 (b) shows a 3D close-up of the red-boxed region in (a), where the detonation front is visualized as the $T = 1600\,\mathrm{K}$ isotherm; the front exhibits a slight inward, concave indentation near the central portion along the $z$-direction, occurring where the equivalence ratio is lowest ($\phi = 0.9$).

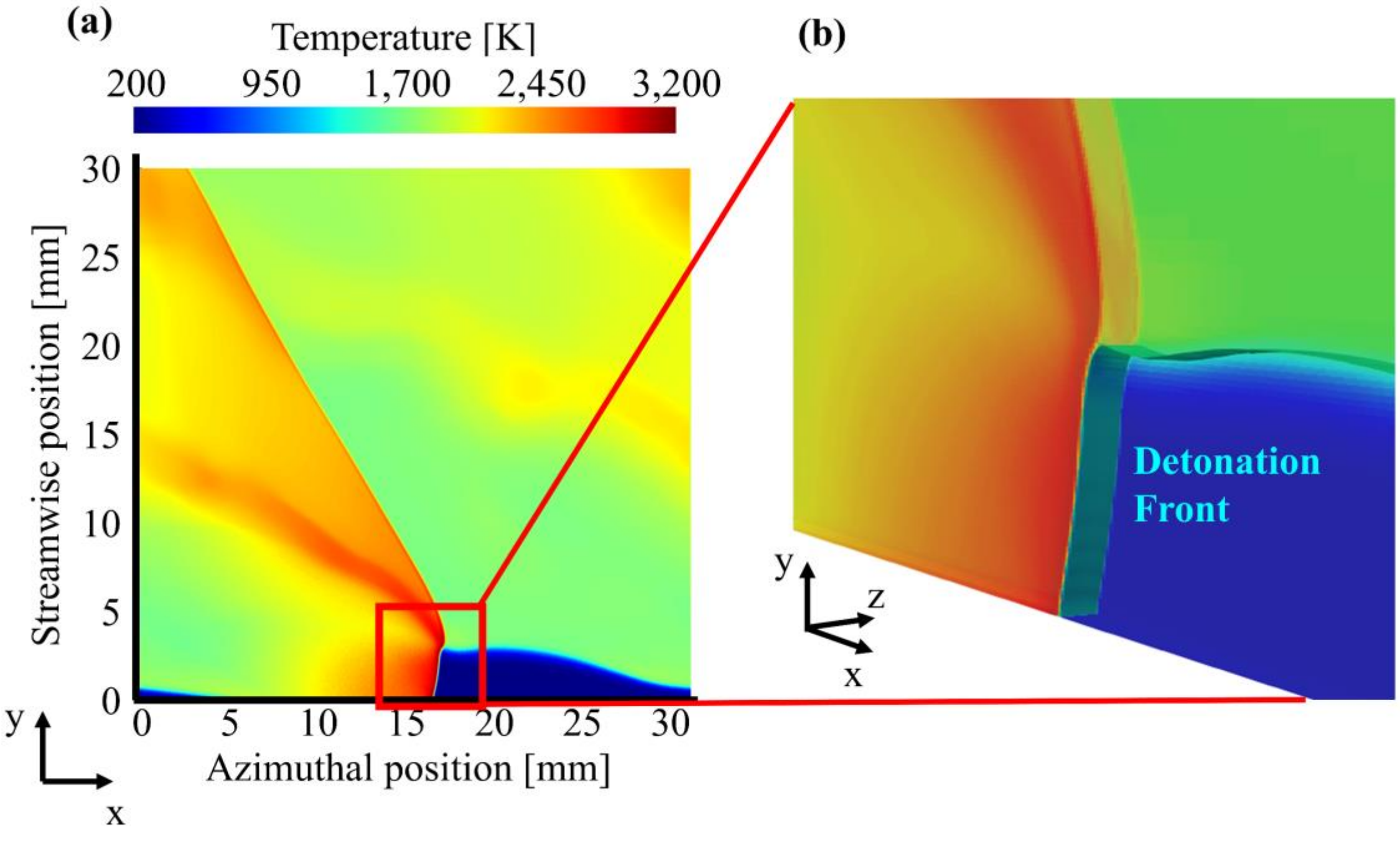

Figure 29. (a) Temperature field in the unwrapped RDE channel $(x, y)$; colors denote temperature (K). (b) 3D magnification of the boxed region; the detonation front is shown as the $T = 1600\,\mathrm{K}$ isotherm.

For clarity, Figure 30 (a) extracts a cross-section of the 3D detonation front by intersecting it with a plane parallel to the $x-z$ plane at $y=1.6$ mm. The intersection curve is then plotted in the $x-z$ plane as streamwise position (vertical axis) versus spanwise position (horizontal axis). This slice highlights the front's slight inward, concave indentation near mid-thickness—occurring where the equivalence ratio is lowest. Figure 30 (b) first shows the theoretical detonation velocity along the spanwise position, obtained by computing the velocity from the locally injected reactants' equivalence ratio shown in Figure 28. This plot explains the shape of Figure 30 (a): the edge regions have higher $\phi$ and thus higher detonation speeds, so they consume more reactants per unit area than the lean interior. To keep pace, the slower interior increases its effective burning area by tilting the front, which enlarges the surface exposed to fresh mixture. Because the outer segments propagate faster, the front tilts inward toward the center, producing the concave indentation seen in Figure 30 (a).

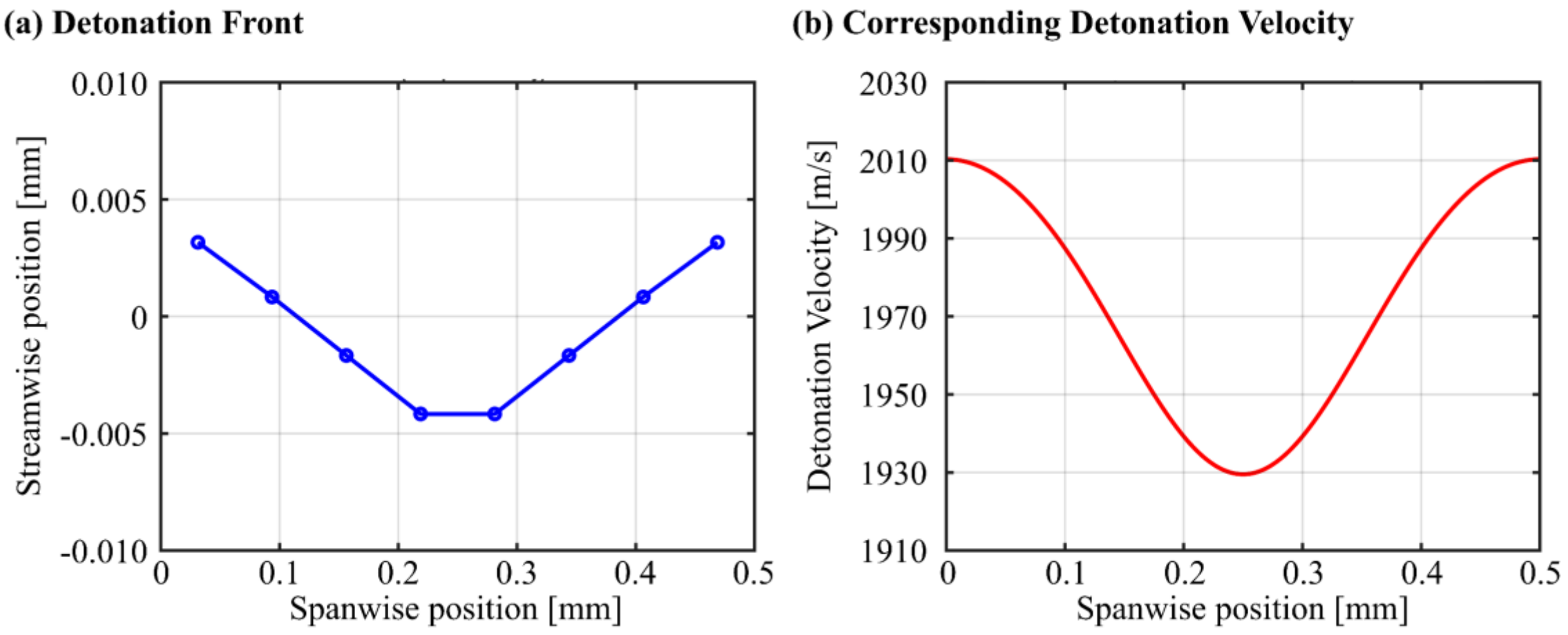

Figure 30. (a) Cross-section of the 3D detonation front from a plane parallel to $x-z$ at $y=1.6$ mm; plotted as streamwise position vs. spanwise position. (b) Theoretical detonation velocity along the spanwise position, computed from the local equivalence ratio.

These results indicate that the solver responds physically to 3D disturbances imposed across the thickness direction: spanwise gradients in the injected equivalence ratio produce the expected coupling between local detonation speed and front geometry, yielding the inward concavity observed in Figure 30 (a) and the velocity distribution in Figure 30 (b). Although the domain does not represent a fully annular three-dimensional RDE, the results provide a preliminary demonstration that the framework can reproduce the qualitative response of the detonation-front geometry to prescribed spanwise mixture non-uniformity. This calculation is intended only as a preliminary qualitative demonstration, and validation using fully three-dimensional annular configurations is required before the quantitative three-dimensional predictive capability of the framework can be established.

*Appendix C. Counter Diffusion Flame Simulation*

To demonstrate the solver's capability for non-premixed combustion, we consider the canonical case of a laminar counterflow diffusion flame. This configuration serves as a fundamental validation problem due to its well-documented experimental database and its strong sensitivity to both transport and chemical-kinetic effects. In this setup, opposing streams of fuel and oxidizer are introduced along the axial direction, establishing a steady, one-dimensional flame stabilized by the balance between convection and diffusion. In the present study, axial and radial profiles of temperature and major species mass fractions are extracted along the centerline and compared directly with available experimental and numerical benchmark data. Such comparisons allow us to assess the solver's accuracy in predicting mixing-controlled, diffusion-limited combustion processes and establish a solid foundation for future applications to more complex non-premixed systems.

Figure 31 summarizes the counterflow configuration used in the simulation. The following conditions follow the experimental conditions by Bufferand et al. [57]. The two-dimensional domain consists of $L_x = 14.1$ mm by $L_y = 12.5$ mm. Fuel enters from the left boundary and oxidizer from the right; the top and bottom boundaries serve as outlets. The fuel stream consists of a mixture of methane and nitrogen with molar compositions of $CH_4$:$N_2$ = 0.103:0.897, and the oxidizer stream consists of oxygen and nitrogen with molar compositions of $O_2$:$N_2$ = 0.773:0.227. Fuel and oxidizer are introduced at 379 K and 340 K, respectively.

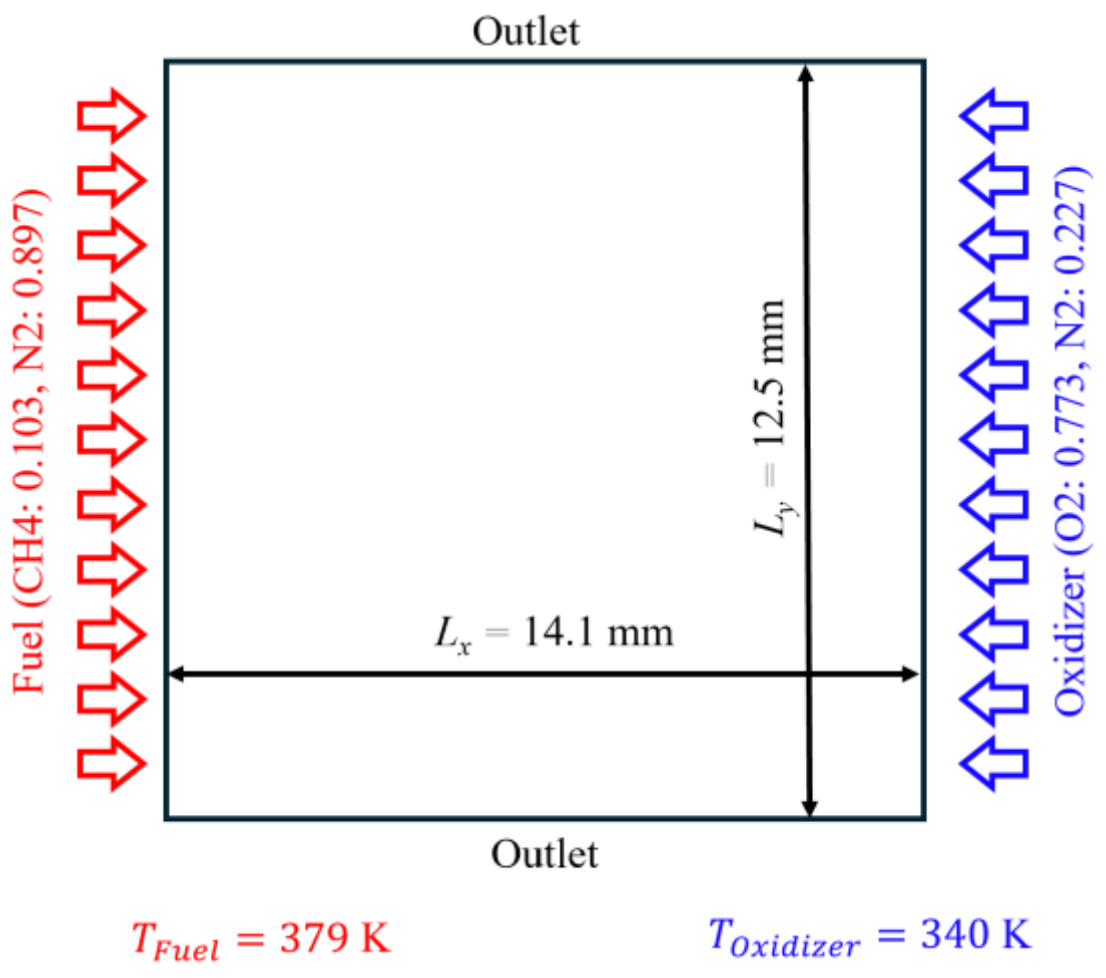

Figure 31. Configuration of numerical setup for a counterflow flame simulation.

Figure 32 (a) plots the centerline temperature as a function of distance from the fuel inlet, following the convention in Bufferand et al. [57]. As in the reference, both the present curve and the reference simulation are shifted in $x$ to enable a like-for-like comparison of flame position. Overall, the present simulation reproduces the experimental trend and peak level well, while the simulated flame zone appears narrower than in the measurements. As discussed by Bufferand et al. [57], this difference arises from using a uniform inlet velocity profile in the simulations, whereas the experiment employs a non-uniform profile; using a uniform profile leads to a thinner flame thickness.

Figure 32 (b) compares centerline mole-fraction profiles of the major species ($CH_4$, $O_2$, CO, $CO_2$) with the measurements and simulations of Bufferand et al. [57]. The simulation reproduces the overall trends—the consumption of CH4 and $O_2$ across the reaction zone and the formation of $CO_2$ and CO—with peak locations and magnitudes broadly consistent with the data. This result indicates that the present solver accurately captures flame structure, species evolution, and mixing characteristics in non-premixed combustion simulations.

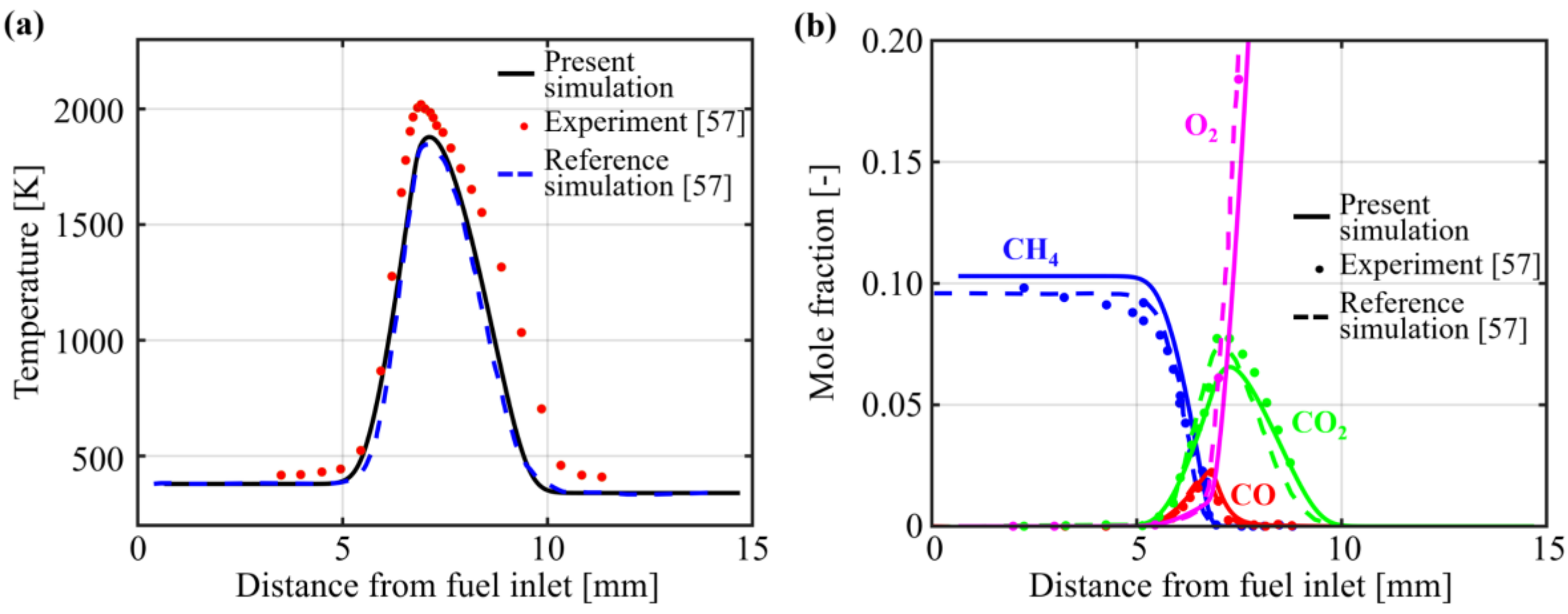

Figure 32. Centerline (a) temperature and (b) mole-fraction profiles ($CH_4$, $O_2$, CO, $CO_2$) vs. distance from the fuel inlet. Present simulation (solid), experiment [57] (dots), and reference simulation [57] (dashed).

## Acknowledgements

This work was supported by "Human Resources Program in Energy Technology" and “International Collaborative Research Center in Energy Technology” of the Korea Institute of Energy Technology Evaluation and Planning (KETEP), granted financial resource from the Ministry of Trade, Industry & Energy, Republic of Korea (Grant #: RS-2021-KP002521). The simulations were carried out on the High-Performance Computing resources at New York University Abu Dhabi (NYUAD).

## Data Availability Statement

The data that support the findings of this study are openly available in GitHub repository *RDE* at https://github.com/kkj9793/RDE.

## Conflict-of-Interest Statement

On behalf of all authors, the corresponding author states that there is no conflict of interest.